\documentclass[a4paper,onecolumn,11pt,accepted=2024-06-16]{quantumarticle}
\pdfoutput=1
\usepackage[utf8]{inputenc}
\usepackage[english]{babel}
\usepackage[T1]{fontenc}
\usepackage{amsmath}
\usepackage{hyperref}

\usepackage{tikz}
\usepackage{lipsum}

\begin{document}

\title{Synthesizing and multiplexing autonomous quantum coherences}

\author{Artur Slobodeniuk}
\email{aslobodeniuk@karlov.mff.cuni.cz}
\affiliation{Department of Optics, Palack\'{y} University, 17. listopadu 12, 77146 Olomouc, Czech Republic}
\affiliation{Department of Condensed Matter Physics, Faculty of Mathematics and Physics,
Charles University, \\ Ke Karlovu 5, CZ-121 16 Prague, Czech Republic}
\orcid{0000-0001-5798-0431}

\author{Tom\'{a}\v{s} Novotn\'{y}}
\email{tno@karlov.mff.cuni.cz}
\affiliation{Department of Condensed Matter Physics, Faculty of Mathematics and Physics,
Charles University, \\ Ke Karlovu 5, CZ-121 16 Prague, Czech Republic}
\orcid{0000-0001-7014-4155}

\author{Radim Filip}
\email{filip@optics.upol.cz}
\affiliation{Department of Optics, Palack\'{y} University, 17. listopadu 12, 77146 Olomouc, Czech Republic}
\orcid{0000-0003-4114-6068}

\maketitle

\begin{abstract}
Quantum coherence is a crucial prerequisite for quantum technologies. Therefore, the robust generation, as autonomous as possible, of quantum coherence remains the essential problem for developing this field. We consider a method of synthesizing and multiplexing quantum coherence from a spin systems without any direct drives only coupled to a bosonic baths. The previous studies in this field have demonstrated that a back-action of the bath to the spin subsystem is important to generate it, however, it simultaneously gives a significant limits to the generated coherence. We propose a viable approach with the bosonic bath that allows overcoming these limits by avoiding the destructive effect of the back-action processes. Using this approach, we suggest advanced synthesis of the quantum coherence non-perturbatively in the spin-boson coupling parameters of multiple bosonic baths to increase and multiplex it for upcoming proof-of-principle experiments.
\end{abstract}

\section{Introduction}

Quantum coherence \cite{Streltsov2017} is a significant and diverse subject of modern quantum physics, phase estimation and thermodynamics 
\cite{Novo2016,Alonso2016,Tan2017,DemkowiczDobrzanski2017,Haase2018,Czajkowski2019,Stella2019,Latune2019,
Francica2020,DelSanto2020,Micadei2020,Diaz2020,Seah2020,Miller2020,Micadei2021,Tupkary2022} 
and a crucial resource of contemporary quantum technology, specifically, quantum metrology 
\cite{Schmitt2017, Zhou2020}, quantum
communication \cite{Reiserer2016, Awschalom2018}, quantum simulators \cite{Hensgens2017, Drost2017, Slot2017},  energy harvesting \cite{Scholes2017, Romero2017}, quantum thermodynamics \cite{Klatzow2019, Ono2020, Latune2021}, 
and quantum computing \cite{Bradley2019, Stephen2019}. A classical, external and strong coherent drive typically generates such quantum coherence as a superposition of energy states. Recently, it has been proposed that there might be a more autonomous alternative, the quantum coherence, from the coupling between a basic system, like a two-level system, and a thermal bath \cite{Guarnieri2018}. It used a composite interaction between the two-level system and thermal bath; in one direction, the incoherent energy of the system pushed the bath coherently, but simultaneously it could receive that coherence of the bath. Both interactions must be present to obtain quantum coherence  in a single two-level system without any external drive, just from a coherent interaction with a bath. 
It is therefore conceptually different from a coherence for a pair of two-level systems from thermal baths
\cite{Brask2015}. It triggered further analysis \cite {Guarnieri2020,Reppert2020,Purkayastha2020,Ancheyta2021, Slobodeniuk2022}; however, it is still not a fully explored phenomenon, without a direct experimental test. For more extensive feasibility, other system-bath topologies generating and detecting more autonomous coherence have to be found.    

Here, we present two crucial steps toward such experimental verifications, considering many separate two-level systems to push the bath coherently, many to receive quantum coherence in parallel and also more baths assisting the process in parallel. Advantageously, we split the single two-level system used in Ref.~\cite{Guarnieri2018} to two separate (drive and output) ones. Using these allowed topologies providing autonomous quantum coherences without a back-action, we propose and study an autonomous synthetisation of coherence from many systems, multiplexing it to many systems and employing jointly different baths to generate the coherences. From the detailed analysis of these cases, we proved a significant result: many systems and baths could be used in parallel to obtain and broadcast autonomous quantum coherences in the experiments.     

The paper is organized as follows. In Sec.~\ref{sec:multispin}, we propose a general method of the calculation of 
spin coherences in systems which contain a bosonic bath interacting with many spins and verify this method 
with the previously obtained results for single spin in 
Refs.~\cite{Purkayastha2020,Ancheyta2021,Slobodeniuk2022}.
In Sec.~\ref{sec:new_method}, we apply the new method to the case of two spins, input and output, interacting separately with the bosonic bath. In Sec.~\ref{sec:synthesizing_coherence}, we extend the previous system to the $M$ input spins and analyze the expression for coherence of single output spin as a function of $M$. 
Then, in Sec.~\ref{sec:coherence_multiplexing} we examine the case of $M$ input and two output spins and consider 
the correlation effects between the output spins. In Sec.~\ref{sec:two_baths}, we explore the case of two bosonic baths coupled separately to $M$ and $N$ input spins, while the output spin is coupled to both baths.
We develop a systematic scheme and calculate the coherence of the output spin in a general form, 
and then discuss the generalization of this problem to larger number of baths and output spins.
Finally, in Sec.~\ref{sec:generalization}, we further generalize our method of calculation by substituting the output spin by an oscillator. In Sec.~\ref{sec:summary}, we summarize our results, discuss the advantages and
limits of the proposed mechanisms of the generation of the cohherence in the spin-bath systems.
Technical details are presented in 4 Appendices.  
%We explicitly consider these terms for the following cases: $M=0$ (to compare 
%with the previously obtained single spin results \cite{Purkayastha2020,Ancheyta2021,Slobodeniuk2022}); 
%to $M=1$ (new driving and output two-spin topology), 
%and finally the extended case of arbitrary number of driving spins for the coherence synthesis $M>1$. 
%Then we extend the considered method for the case of $M+2$ spins for the coherence multiplication 
%and calculate the 2-spin reduced density operator. 
%Finally, we generalize the previous problem by extending the number of the independent bosonic baths in the system
%to obtain coherence through them.  
%We consider the simplest case of 2 different baths and $M+N+1$ spins. The first bath is coupled to $M$ driving spins, 
%the second bath is coupled to $N$ driving spins, and the remaining outputspin is coupled to both baths. 
%\section{Spin-boson model}

\section{Multi-spin interaction with thermal bath}
\label{sec:multispin}

First, we modify the model considered in the Refs.~\cite{Purkayastha2020,Ancheyta2021,Slobodeniuk2022} to open more possibilities for the synthesization and multiplexing by separating a single two-level system into driving and receiving two-level systems.    

We consider the Hamiltonian of the system $H=H_B+H_S+H_{SB}=H_0+H_{SB}$. Here 
\begin{equation}
H_B=\sum_k \Omega_k b_k^\dag b_k,
\end{equation}
is the Hamiltonian of bosonic excitations with the spectrum $\Omega_k>0$. 
The operators $b_k, b_k^\dag$ satisfy the canonical commutation relations $[b_k, b_q^\dag]=\delta_{kq}$, 
where $\delta_{kq}$ is the Kronecker symbol. Furthermore, 
\begin{equation}
H_S=\sum_{j=1}^{M+1}\frac{\omega_j}{2}\sigma_j^z,
\end{equation} 
with $\omega_j>0$ is the Hamiltonian of $M+1$ spins, $M$ driving the bath to get coherence 
there and the $(M+1)$th receiving the coherence. Finally,
\begin{align}
\label{eq:spin_boson}
H_{SB}=&\sum_{j=1}^{M+1}\boldsymbol{\sigma}_j\cdot\boldsymbol{n}_j
\sum_k \lambda_k(b_k^\dag+b_k),
\end{align} 
describes the interaction of the spins with the bosonic system. Here, we have introduced the notation
$\boldsymbol{\sigma}\cdot\boldsymbol{n}\equiv\sigma^x n^x+\sigma^y n^y+\sigma^z n^z$, with the Pauli matrices 
$\sigma^x,\sigma^y,\sigma^z$ and vector of the coupling strength parameters $\boldsymbol{n}=(n^x,n^y,n^z)$. 

We are interested in the reduced density matrix of the spin system, which can be obtained from 
the full canonical density matrix $\rho\equiv e^{-\beta H}/Z$ by tracing out over the bosonic degrees of freedom
\begin{equation}
\rho_S=Z^{-1}\text{Tr}_B\Big[e^{-\beta H}\Big],
\end{equation}
where $Z$ is the partition function of the full system 
\begin{equation}
Z=\text{Tr}_{S}\Big[\text{Tr}_B\Big[e^{-\beta H}\Big]\Big].
\end{equation}
We evaluate the reduced density matrix using the following method. 
First we present the operator exponent in the form 
\begin{equation}
e^{-\beta H}=e^{-\beta H_0}\Big(e^{\beta H_0}e^{-\beta H}\Big)=e^{-\beta H_0}U(\beta).
\end{equation}
The operator $U(\tau)$ satisfies the differential equation in the domain $\tau\in[0,\beta]$
with the initial condition $U(0)=1$
\begin{equation}
\frac{\partial U(\tau)}{\partial\tau}=-\widetilde{H}_{SB}(\tau)U(\tau),
\end{equation}
where $\widetilde{H}_{SB}(\tau)\equiv e^{\tau H_0}H_{SB}e^{-\tau H_0}$. 
The solution of the equation can be presented in the form of the chronologically ordered 
(in the imaginary Matsubara time $\tau$) exponent
\begin{align}
\label{eq:t_ordered}
U(\beta)=&T_\tau\Big\{e^{-\int_0^\beta d\tau \widetilde{H}_{SB}(\tau)}\Big\}=
1-\int_0^\beta d\tau_1 \widetilde{H}_{SB}(\tau_1) +
\nonumber \\ +& \int_0^\beta d\tau_1 \widetilde{H}_{SB}(\tau_1)
\int_0^{\tau_1} d\tau_2 \widetilde{H}_{SB}(\tau_2) + \dots.  
\end{align} 
Using this result we rewrite the reduced density matrix as 
\begin{equation}
\label{eq:reduced_density_operator}
\rho_S=\frac{e^{-\beta H_S}}{Z_S}
\frac{\Big\langle T_\tau\Big\{e^{-\int_0^\beta d\tau \widetilde{H}_{SB}(\tau)}\Big\}\Big\rangle_B}
{\Big\langle\Big\langle T_\tau\Big\{e^{-\int_0^\beta d\tau \widetilde{H}_{SB}(\tau)}\Big\}\Big\rangle_B\Big\rangle_S}. 
\end{equation}
Here the averaging procedures over the spin $(S)$ and bosonic $(B)$ degrees of freedom read   
\begin{align}
\langle\star\rangle_S\equiv&Z_S^{-1}\text{Tr}_S\Big[e^{-\beta H_S}\star\Big], \\ 
\langle\star\rangle_B\equiv&Z_B^{-1}\text{Tr}_B\Big[e^{-\beta H_B}\star\Big].
\end{align}
with
\begin{equation}
Z_S\equiv\text{Tr}_S\Big[e^{-\beta H_S}\Big], \quad Z_B\equiv\text{Tr}_B\Big[e^{-\beta H_B}\Big],
\end{equation}
being the partition functions of the free spin $(Z_S)$ and boson $(Z_B)$ subsystems, respectively. 
Note that the numerator of Eq.~(\ref{eq:reduced_density_operator}) can be presented as $e^{-\beta H_{eff}}$, 
the denominator then transforms into $\text{Tr}_S[e^{-\beta H_{eff}}]$, see details in Ref.~\cite{Lobeiko2022}.
However, in the current study we use another way by evaluating perturbatively the $T_\tau$-ordered exponent. 

The reduced density operator $\rho_S$ depends on the $M+1$ spin degrees of freedom. Our idea is based on the 
observation that the bosonic bath effectively couples the spin degrees of freedom. 
Part of these coupling terms is responsible for the generation of a non-zero value of the 
$\sigma^x_{M+1}$-operator (coherence), which can be calculated as 
\begin{equation}
\langle \sigma^x_{M+1}\rangle\equiv\text{Tr}_S[\rho_S\sigma^x_{M+1}].
\end{equation}  

The aforementioned spin-spin interaction terms correspond to the leading terms of the perturbation series for 
the reduced density matrix $\rho_S$ in the spin-boson coupling parameters and can be deduced from the following expression in the 
weak-coupling regime (see derivation in Appendix~\ref{app:derivation})
\begin{align}
\Big\langle T_\tau\Big\{&e^{-\int_0^\beta d\tau \widetilde{H}_{SB}(\tau)}\Big\}\Big\rangle_B
\approx 1+\int_0^\infty d\xi\,\mathcal{I}(\xi)\times \nonumber \\
\times&\int_0^\beta d\tau \int_0^\tau d\tau' 
\phi(\xi,\tau-\tau')F(\tau)F(\tau').
\end{align}
Here, we have introduced the bosonic spectral density function 
\begin{equation}
\mathcal{I}(\xi)\equiv\sum_k \lambda_k^2\delta(\xi-\Omega_k),
\end{equation} 
the function
\begin{equation}
\phi(\xi,\tau-\tau')=\frac{e^{(\tau-\tau')\xi}}{e^{\beta\xi}-1} +
\frac{e^{-(\tau-\tau')\xi}}{1-e^{-\beta\xi}},
\end{equation} 
and $\tau$-dependent multi-spin operator
\begin{equation}
\label{eq:F_tau}
F(\tau)=e^{\tau H_S}\Big[\sum_{j=1}^{M+1}\boldsymbol{\sigma}_j\cdot\boldsymbol{n}_j\Big]e^{-\tau H_S}.
\end{equation}
\begin{figure*}
\centering
\includegraphics[width=\linewidth]{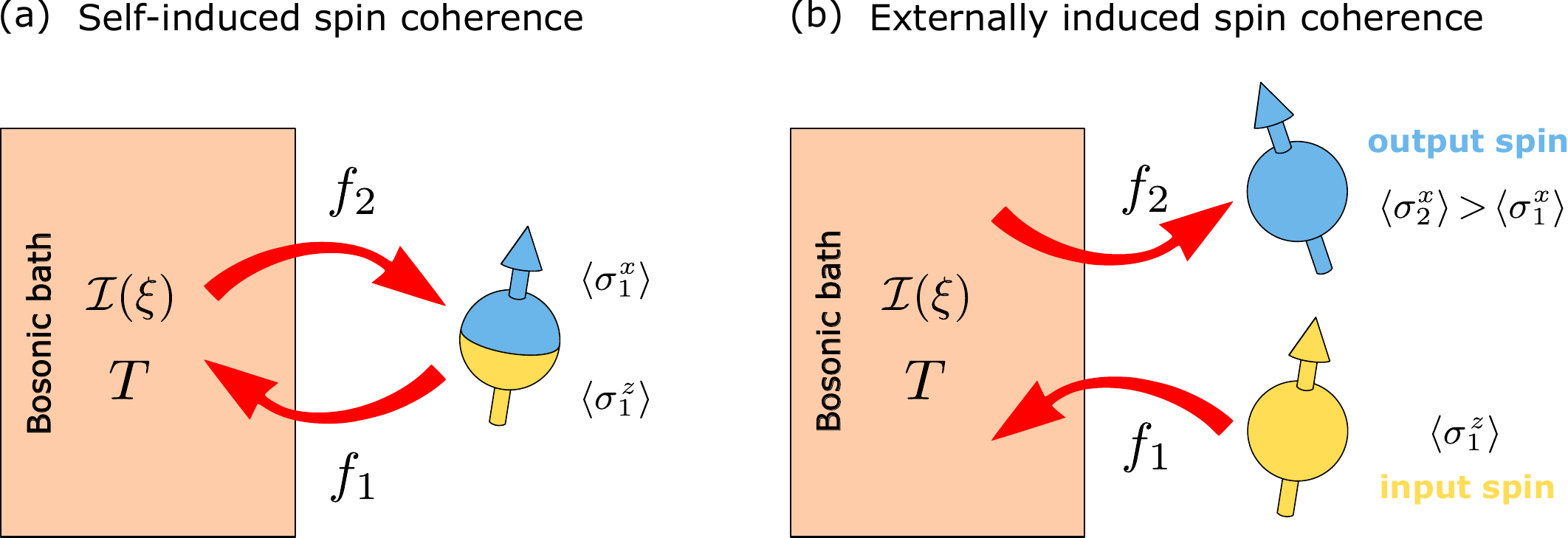}
\caption{\label{fig:fig_1} Generation of the spin coherence   
        in two different schemes of the spin-bath interaction. (a) Self-induced spin coherence method from
				Refs.~\cite{Slobodeniuk2022,Ancheyta2021,Purkayastha2020}.
				The single spin ``polarizes'' the bosonic 
				subsystem with spectral density function $\mathcal{I}(\xi)$ at temperature $T$ via the $f_1\sigma_1^z$ term. 
				The polarized bosonic bath generates indirectly the coherence in the
				{\it same} spin $\langle\sigma_1^x\rangle$ via the $f_2\sigma_1^x$ term, i.e., as the {\it back reaction} to 
				the spin system from the bosonic bath. (b) Externally induced spin coherence method. 
				The first (input) spin polarizes the bosonic bath 
				via the $f_1\sigma_1^z$ term.
				The polarized bosonic bath generates the coherence in the {\it output} spin 
				$\langle\sigma_2^x\rangle$ via the $f_2\sigma_2^x$ term, i.e., 
				as the {\it transfer of spin coherence through the bath} to the second spin 
				system.}
\end{figure*} 

For a comparison, we first reconsider the case with the following single-spin Hamiltonian
\cite{Slobodeniuk2022,Ancheyta2021,Purkayastha2020}
\begin{equation}
%H_S=\frac{\omega_1}{2}\sigma_1^z, \quad 
H_{BS}=[f_1\sigma_1^z+f_2\sigma_1^x]\sum_{k}\lambda_k (b_k+b_k^\dag).
\end{equation} 
It corresponds to the situation of one spin coupled to the bosonic thermal bath simultaneously via the
$\sigma^z$ and $\sigma^x$ spin operators, i.e., $\boldsymbol{n}=(f_2,0,f_1)$. The corresponding system is depicted in Fig.~\ref{fig:fig_1} a).

Using the result of Appendix~\ref{app:derivation}, we can write the leading spin-spin correlation term 
of the reduced spin density operator in this case
\begin{widetext}
\begin{align}
\label{eq:FF}
F(\tau)F(\tau')=\,&[f_1\sigma_1^z+f_2\sigma_1^x(\tau)][f_1\sigma_1^z+f_2\sigma_1^x(\tau')]=\nonumber \\=&
%f_1^2+f_1f_2[\sigma_1^z\sigma_1^x(\tau')+\sigma_1^x(\tau)\sigma_1^z]+
%f_2^2\sigma_1^x(\tau)\sigma_1^x(\tau')=\nonumber 
f_1^2+f_2^2\cosh(\omega_1(\tau-\tau'))+
f_2^2\sinh(\omega_1(\tau-\tau'))\sigma_1^z+\nonumber \\+&
f_1f_2[\sinh(\omega_1\tau')-\sinh(\omega_1\tau)]\sigma_1^x+
if_1f_2[\cosh(\omega_1\tau')-\cosh(\omega_1\tau)]\sigma_1^y, 
\end{align} 
\end{widetext}
In the first line of Eq.~(\ref{eq:FF}) we introduced the $\tau$-dependent operators
\begin{equation}
\sigma^j(\tau)\equiv e^{\tau\frac{\omega}{2}\sigma^z} \sigma^j e^{-\tau\frac{\omega}{2}\sigma^z}, 
\end{equation} 
with $j=x,y,z$. Using the algebra of Pauli matrices 
$\sigma_j\sigma_k=\delta_{jk}\mathcal{I}+i\varepsilon_{jkl}\sigma_l$, 
where $\mathcal{I}$ is a $2\times2$ unit matrix and $\varepsilon_{jkl}$ is the Levi-Civita symbol, 
we evaluated 
$\sigma^x(\tau)=\cosh(\omega\tau)\sigma^x+i\sinh(\omega\tau)\sigma^y$, 
$\sigma^y(\tau)=\cosh(\omega\tau)\sigma^y-i\sinh(\omega\tau)\sigma^x$, 
$\sigma^z(\tau)=\sigma^z$, 
and then calculated the coherence $\langle\sigma_1^x\rangle$ 
in the leading order of the perturbation theory  
\begin{widetext}
\begin{align}
\langle\sigma_1^x\rangle\equiv&\text{Tr}_S\Big[\rho_S\sigma_1^x\Big]\approx 
\int_0^\infty d\xi\,\mathcal{I}(\xi)\int_0^\beta d\tau \int_0^\tau d\tau' 
\phi(\xi,\tau-\tau')\langle F(\tau)F(\tau')\sigma_1^x\rangle_S=\nonumber \\
%f_1f_2\int_0^\infty d\xi\,\mathcal{I}(\xi) \times\nonumber \\
%\times &
%\int_0^\beta d\tau \int_0^\tau d\tau' 
%\phi(\tau-\tau')\Big( [\sinh(\omega_1\tau')-\sinh(\omega_1\tau)]
%-\tanh\Big(\frac{\beta\omega_1}{2}\Big)[\cosh(\omega_1\tau')-\cosh(\omega_1\tau)]\Big)=
=&-4f_1f_2\tanh\left(\frac{\beta\omega_1}{2}\right)\int_0^\infty d\xi\,\mathcal{I}(\xi) 
\frac{\xi\coth\left(\frac{\beta\xi}{2}\right)-\omega_1\coth\left(\frac{\beta\omega_1}{2}\right)}
{\xi\left(\xi^2-\omega_1^2\right)}.
\end{align} 
\end{widetext}
This answer coincides with the previously obtained one \cite{Purkayastha2020,Ancheyta2021,Slobodeniuk2022}.
Note that the complex structure of the integrand is a result of the dynamical 
back reaction of the bosonic bath onto the spin system. 
This is a consequence of the special coupling of the spin system to the bosonic bath, 
containing both the $\sigma^x$ and $\sigma^z$ coupling terms. 
It allows self-induced coherence through the bath by the spin itself but also 
creates dynamical terms that limit the amount of coherence.  
In order to eliminate these dynamical terms we introduce two groups of spin systems, 
where the spins of the first (second) group interact with the bosonic bath 
only via the $\sigma^z$ ($\sigma^x$) coupling term. 
In this configuration the first group of spins influences the bosonic system as driving spins and then
the affected bosonic bath generates the nonzero coherence in the second group of output spins. 
The simplest case of such systems is considered in the next section.   
  
\section{New two-spin method $(M=1)$}
\label{sec:new_method}

For further development and comparison, we propose the basic case 
$\boldsymbol{n}_1=(0,0,f_1)$, $\boldsymbol{n}_2=(f_2,0,0)$,  
with $f_1,f_2\in\textrm{R}$. The corresponding system is depicted in Fig.~\ref{fig:fig_1} b). 
For this case the spin-spin correlation term reads
\begin{widetext}
\begin{align}
F(\tau)F(\tau')=&[f_1\sigma_1^z+f_2\sigma_2^x(\tau)][f_1\sigma_1^z+f_2\sigma_2^x(\tau')]=\nonumber \\=&
%=f_1^2+f_1f_2[\sigma_1^z\sigma_2^x(\tau')+\sigma_2^x(\tau)\sigma_1^z]+
%f_2^2\sigma_2^x(\tau)\sigma_2^x(\tau')=\nonumber 
%\\=&
f_1^2+f_2^2\cosh(\omega_2(\tau-\tau'))+f_2^2\sinh(\omega_2(\tau-\tau'))\sigma_2^z +\nonumber \\
+&f_1f_2\Big([\cosh(\omega_2\tau')+\cosh(\omega_2\tau)]\sigma_1^z\sigma_2^x+
i[\sinh(\omega_2\tau')+\sinh(\omega_2\tau)]\sigma_1^z\sigma_2^y\Big).
\end{align}
\end{widetext}
Using this result we obtain 
\begin{align}
\label{eq:second_spin_coherence}
\langle\sigma_2^x\rangle\equiv\text{Tr}_S\Big[\rho_S\sigma_2^x\Big]\approx 
%\int_0^\infty d\xi\,\mathcal{I}(\xi)\int_0^\beta d\tau \int_0^\tau d\tau' 
%\phi(\tau-\tau')\langle F(\tau)F(\tau')\sigma_2^x\rangle_S=
%-f_1f_2\tanh\Big(\frac{\beta\omega_1}{2}\Big)\int_0^\infty d\xi\,\mathcal{I}(\xi) \times\nonumber \\
%\times &
%\int_0^\beta d\tau \int_0^\tau d\tau' 
%\phi(\tau-\tau')\Big( 
% [\cosh(\omega_2\tau')+\cosh(\omega_2\tau)]-\tanh\Big(\frac{\beta\omega_2}{2}\Big)
% [\sinh(\omega_2\tau')+\sinh(\omega_2\tau)]\Big)=
%\nonumber \\=&
% -4f_1f_2\tanh\Big(\frac{\beta\omega_1}{2}\Big)\frac{\tanh\left(\frac{\beta\omega_2}{2}\right)}{\omega_2}
%\int_0^\infty d\xi\,\frac{\mathcal{I}(\xi)}{\xi}=
%\nonumber \\=&
-4f_1f_2\tanh\Big(\frac{\beta\omega_1}{2}\Big)\frac{\tanh\left(\frac{\beta\omega_2}{2}\right)}{\omega_2}\Omega, 
\end{align}
where we have introduced the quantity $\Omega\equiv\int_0^\infty d\xi\,\mathcal{I}(\xi)/\xi$, 
with meaning of the reorganization energy of the bosonic bath. 
Note that this result coincides with the mean field (static) result for the original case in the low-temperature limit \cite{Slobodeniuk2022}. 

Let us compare the results: the original self-induced method \cite{Slobodeniuk2022,Purkayastha2020}, where a single spin was both the driving and output system, 
and the new method splitting these roles to separate spins. 
To do it we first put $\omega_2=\omega_1=\omega$ in the new method to discuss a resonant case
and rewrite both expressions in the following form
\begin{widetext}
\begin{align}
\langle\sigma_1^x\rangle=&-4f_1f_2\tanh\left(\frac{\beta\omega}{2}\right)\int_0^\infty d\xi\,
\frac{\mathcal{I}(\xi)}{\xi} 
\frac{\xi\coth\left(\frac{\beta\xi}{2}\right)-\omega\coth\left(\frac{\beta\omega}{2}\right)}
{\left(\xi^2-\omega^2\right)}= \nonumber \\ &=-\int_0^\infty d\xi\,
\frac{\mathcal{I}(\xi)}{\xi}\mathcal{F}_1(\beta,\omega,\xi),\\
\langle\sigma_2^x\rangle=&-4f_1f_2\tanh\Big(\frac{\beta\omega}{2}\Big)
\int_0^\infty d\xi\,\frac{\mathcal{I}(\xi)}{\xi}\frac{\tanh\left(\frac{\beta\omega}{2}\right)}{\omega}=-
\int_0^\infty d\xi\,\frac{\mathcal{I}(\xi)}{\xi}\mathcal{F}_2(\beta,\omega,\xi). 
\end{align}
\end{widetext}
Taking into account the positivity of $\mathcal{I}(\xi)>0$ and the fact that 
$\mathcal{F}_2(\beta,\omega,\xi)>\mathcal{F}_1(\beta,\omega,\xi)>0$ for any value of $\beta$ and $\omega$ 
we obtain that $|\langle\sigma_2^x\rangle|>|\langle\sigma_1^x\rangle|$ for the resonant case. 
It demonstrates that the new method of 
generation of the coherence using two spins with distributed roles 
is more effective than the previously proposed method \cite{Purkayastha2020}, where a single spin had to play double role, both to coherently displace the bath by its thermal population and receive that coherence back. 

\begin{figure*}[t]
	\centering
	\includegraphics[width=\linewidth]{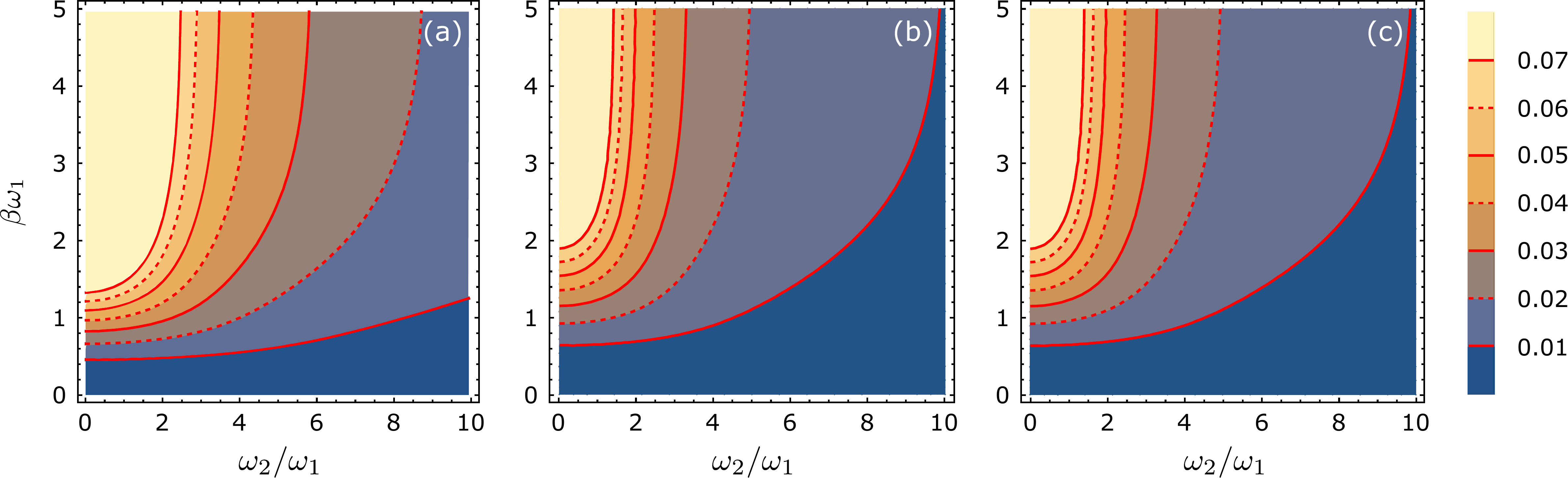} 
\caption{Density plot of normalized coherence $\langle\sigma_2^x\rangle/(-4f_1f_2\lambda)$ 
        for (a) sub-Ohmic with $s=0.5$, (b) Ohmic with $s=1$, and (c) super-Ohmic with $s=2$ cases
				in the domain $\omega_2/\omega_1 \in[0,10]$ and $\beta\omega_1 \in[0,5]$. 
				One can observe that the results for $s=1$ and $s=2$ coincide, which reflects the fact that 
				$\langle\sigma_2^x\rangle\propto \Gamma(s)$.}	
\label{fig:picture_sigma_2}
\end{figure*}

In order to know the effectiveness of the generation of the coherence with the new method for a couple 
of spins we consider a general case of different frequencies $\omega_1\neq \omega_2$ with   
the canonical spectral density function 
\begin{equation}
\label{eq:spectral}
\mathcal{I}(\xi)=\lambda \frac{\xi^s}{\omega_c^{s-1}}e^{-\xi/\omega_c}.
\end{equation} 
The dimensionless parameter $\lambda$ describes the strength of the spectral density function, while $\omega_c$ represents the energy cut-off \cite{Guarnieri2018,Weiss2012,Leggett1987}.     
Note that for this spectral density the coherence $\langle\sigma_2^x\rangle$ (\ref{eq:second_spin_coherence}) 
can be calculated analytically 
\begin{align}
\langle\sigma_2^x\rangle=&-4f_1f_2\tanh\Big(\frac{\beta\omega_1}{2}\Big)
\frac{\tanh\left(\frac{\beta\omega_2}{2}\right)}{\omega_2}\lambda\omega_c\Gamma(s), 
\end{align} 
where $\Gamma(s)$ is the Gamma function. The normalized coherence $\langle\sigma_2^x\rangle/(-4f_1f_2\lambda)$ 
for different values of the parameter $s=0.5;1;2$ as a function of the dimensionless parameters  $\beta\omega_1$
and $\omega_2/\omega_1$ is presented in Fig.~\ref{fig:picture_sigma_2}. 
The coherence linearly growths with the strength $\lambda$ of the spectral density function and 
its cut-off energy $\omega_c$. 
%The answer also demonstrates that for fixed values 
%of $\lambda$ and $\omega_c$ the larger coherence can be obtained in the domains $0<s<1$ or $s>2$. 
As a function of the temperature $T=1/\beta$ the coherence is maximized 
in the $T\rightarrow 0$ limit 
\begin{align}
\langle\sigma_2^x\rangle|_{T=0}=&-4f_1f_2\lambda\frac{\omega_c}{\omega_2}\Gamma(s). 
\end{align}   
Finally, the generated coherence for the fixed temperature $T=1/\beta$ can be increased by 
increasing the frequency of the first spin $\omega_1$ and decreasing 
the frequency of the second spin $\omega_2$. However, 
the limit $\omega_2\rightarrow 0$ can't be taken since it will violate 
the conditions of the perturbation theory. Namely, the aforementioned perturbation 
analysis is applicable as long as the condition $\omega_2\gg 4f_1f_2\Omega$ is satisfied, see details in 
Appendix~\ref{app:sigma_x}. The generalized non-perturbative analysis of 
the case with arbitrary $\omega_2$ is presented later in Sec.~\ref{sec:two_baths}.  
\begin{figure*}[t]
	\centering
	\includegraphics[width=\linewidth]{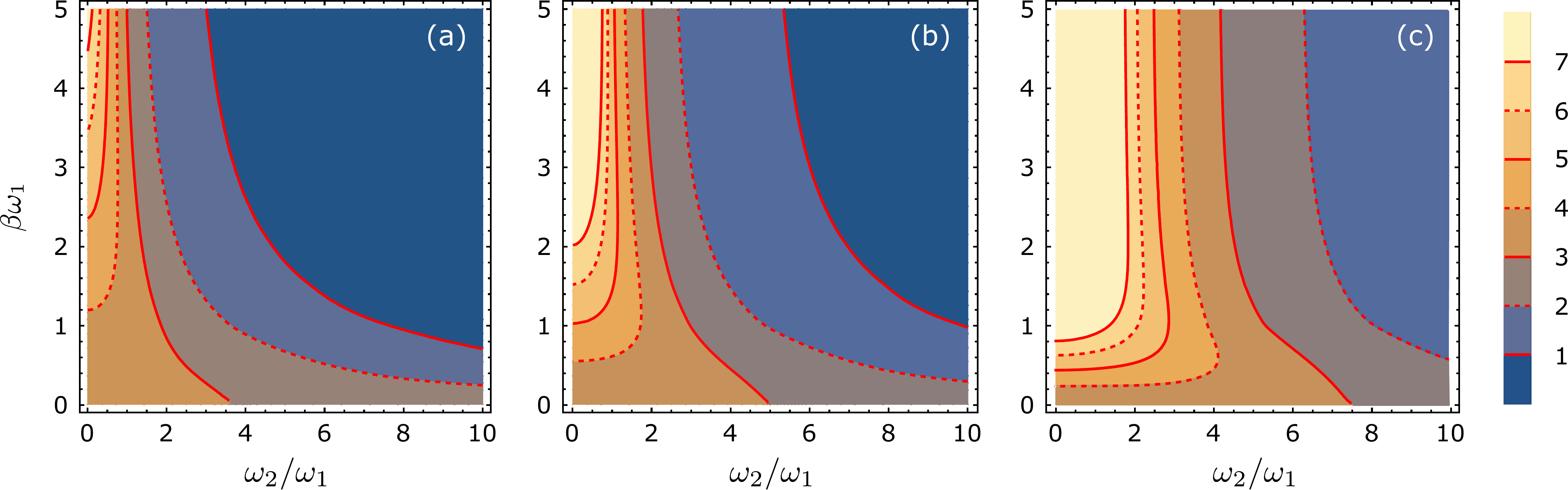} 
\caption{Density plot of $\langle\sigma_2^x\rangle/\langle\sigma_1^x\rangle$ 
        for (a) sub-Ohmic with $s=0.5$, (b) Ohmic with $s=1$, and (c) super-Ohmic with $s=2$ cases
				in the domain $\omega_2/\omega_1 \in[0,10]$ and $\beta\omega_1 \in[0,5]$. 
				One can observe that the new method of generation of the
				coherence becomes ineffective comparing to the previous one, i.e., 
				$\langle\sigma_2^x\rangle<\langle\sigma_1^x\rangle$, only for relatively large values of the parameter 
				$\omega_2/\omega_1$.}
\label{fig:picture_prx}				
\end{figure*}

Note that $\langle\sigma_1^x\rangle$ and $\langle\sigma_2^x\rangle$ are non-monotonic functions of the parametr $s$.
Therefore it is not obvious for which parameters $\omega_2$, $\beta$ and $s$ the coherence $\langle\sigma_2^x\rangle$ 
is larger than the coherence $\langle\sigma_1^x\rangle$. 
In order to clarify this issue we calculate the ratio $\langle\sigma_2^x\rangle/\langle\sigma_1^x\rangle$ for
the dimensionless parameters $\omega_2/\omega_1$ and $\beta\omega_1$ 
\begin{widetext}
\begin{equation}
\frac{\langle\sigma_2^x\rangle}{\langle\sigma_1^x\rangle}=
\Big(\frac{\omega_c}{\omega_1}\Big)^s\frac{\Gamma(s)\tanh\Big(\frac{\beta \omega_1}{2}\frac{\omega_2}{\omega_1}\Big)}
{(\omega_2/\omega_1)}
\left[\int_0^\infty dx x^{s-1} e^{-\omega_1 x/\omega_c}\frac{x\coth\Big(\frac{\beta \omega_1}{2}
\frac{\omega_2}{\omega_1}\Big)-\coth\Big(\frac{\beta\omega_1}{2}\Big)}{x^2-1}\right]^{-1},
\end{equation}
\end{widetext}
and for three values of the parameter $s=0.5;1;2$. 
The corresponding plots for the case $\omega_1/\omega_c=0.1$ are presented in Fig.~\ref{fig:picture_prx}.
The plots demonstrate that the new method of generation becomes effective for small ratio $\omega_2/\omega_1$. 
It can be understood from analysis of the expression for the coherence of the second spin 
Eq.~(\ref{eq:second_spin_coherence})
\begin{align}
\langle\sigma_2^x\rangle=&-4f_1f_2\tanh\Big(\frac{\beta\omega_1}{2}\Big)
\frac{\tanh\left(\frac{\beta\omega_2}{2}\right)}{\omega_2}\Omega. 
\end{align} 
As one can see the larger $\omega_1$ corresponds to the larger value of 
$0<\tanh(\beta\omega_1/2)<1$. From other side the multiplier $\tanh(\beta\omega_2/2)/\omega_2$ as a function 
of $\omega_2$ is a decreasing function. Therefore, the larger value of this multiplier can be reached at small values of $\omega_2$. 

\section{Synthesizing coherence from $M$ spins through a single bath}
\label{sec:synthesizing_coherence}

Due to the new method we consider we can now address the first general problem,
if the coherence can be synthesized from many driving spins, in general, 
with specific different couplings and frequencies. 

Let us consider the case $\boldsymbol{n}_j=(0,0,f_1)$, with $j=1,2,\dots M$, 
$\boldsymbol{n}_{M+1}=(f_2,0,0)$. Then we have 
\begin{widetext}
\begin{align}
F(\tau)F(\tau')=&%\Big[f_1\sum_{j=1}^M\sigma_j^z+f_2\sigma_{M+1}^x(\tau)\Big]
%\Big[f_1\sum_{j=1}^M\sigma_j^z+f_2\sigma_{M+1}^x(\tau')\Big]=
f_1f_2\sum_{j=1}^M[\cosh(\omega_{M+1}\tau')+\cosh(\omega_{M+1}\tau)]\sigma_j^z\sigma_{M+1}^x+\nonumber \\+& 
if_1f_2\sum_{j=1}^M[\sinh(\omega_{M+1}\tau')+\sinh(\omega_{M+1}\tau)]\sigma_j^z\sigma_{M+1}^y+
\nonumber \\ +&
%=&f_1^2\sum_{j,l=1}^M\sigma_j^z\sigma_l^z+f_1f_2\sum_{j=1}^M [\sigma_j^z\sigma_{M+1}^x(\tau')+
%\sigma_{M+1}^x(\tau)\sigma_j^z]+f_2^2\sigma_{M+1}^x(\tau)\sigma_{M+1}^x(\tau')=\nonumber 
f_1^2\sum_{j,l=1}^M\sigma_j^z\sigma_l^z+f_2^2\cosh(\omega_{M+1}(\tau-\tau'))+f_2^2\sinh(\omega_{M+1}(\tau-\tau'))
\sigma_{M+1}^z.
\end{align}
\end{widetext}
Using this result we calculate $\langle\sigma_{M+1}^x\rangle=\text{Tr}_S[\rho_S\sigma_{M+1}^x]$
\begin{align}
\label{eq:multispin_result}
\langle\sigma_{M+1}^x\rangle\approx 
%\int_0^\infty d\xi\,\mathcal{I}(\xi)\int_0^\beta d\tau \int_0^\tau d\tau' 
%\phi(\tau-\tau')\langle F(\tau)F(\tau')\sigma_{M+1}^x\rangle_S=
%-f_1f_2\sum_{j=1}^M\tanh\Big(\frac{\beta\omega_j}{2}\Big) 
%\int_0^\infty d\xi\,\mathcal{I}(\xi) 
%\times\nonumber \\ \times&
%\int_0^\beta d\tau \int_0^\tau d\tau' 
%\phi(\tau-\tau')\Big( 
% [\cosh(\omega_{M+1}\tau')+\cosh(\omega_{M+1}\tau)]-\tanh\Big(\frac{\beta\omega_{M+1}}{2}\Big)
% [\sinh(\omega_{M+1}\tau')+\sinh(\omega_{M+1}\tau)]\Big)=
%\nonumber \\=&
 -4f_1f_2\Big[\sum_{j=1}^M\tanh\Big(\frac{\beta\omega_j}{2}\Big)\Big]
\frac{\tanh\left(\frac{\beta\omega_{M+1}}{2}\right)}{\omega_{M+1}}\Omega.
\end{align}
Comparing this result with the previous case (\ref{eq:second_spin_coherence}) one concludes that each of $M$ spins 
contributes cumulatively to the $(M+1)$th spin coherence  
$\langle\sigma_{M+1}^x\rangle \propto \sum_{j=1}^M \langle\sigma_j^z\rangle_S$.
It means, the bath can equally accumulate the coherence from the same driving spins 
and then uses it to make the output spin 
equivalently coherent even if the coupling $f_1f_2$ in Eq.~(\ref{eq:multispin_result}) decreases $M$ times.  

This result can be generalized for the case of $M$ spins with different couplings to the bath system, 
i.e., for the case  
\begin{align}
H_{SB}=\Big[\sum_{j=1}^M \boldsymbol{\sigma}_j\cdot\boldsymbol{n}_j+ 
\boldsymbol{\sigma}_{M+1}\cdot\boldsymbol{n}_{M+1}\Big]
\sigma_k \lambda_k(b_k^\dag+b_k), 
\end{align}
with different $\boldsymbol{n}_j=(0,0,f_1^{(j)})$, for $j=1,2,\dots M$ and $\boldsymbol{n}_{M+1}=(f_2,0,0)$.
Repeating the previous calculations for this case we obtain   
\begin{align}
\langle\sigma_{M+1}^x\rangle\approx 
 -4f_2\Big[\sum_{j=1}^M f_1^{(j)}\tanh\Big(\frac{\beta\omega_j}{2}\Big)\Big]
\frac{\tanh\left(\frac{\beta\omega_{M+1}}{2}\right)}{\omega_{M+1}}\Omega.
\end{align}
It is convenient to introduce the density function of the coupling parameters 
\begin{align}
f_1(\omega)\equiv\sum_{j=1}^M \delta(\omega-\omega_j)f_1^{(j)}, 
\end{align}
and rewrite the expression for the coherence in the form 
\begin{align}
\langle\sigma_{M+1}^x\rangle\approx 
 -4f_2 \Big[\int_{-\infty}^\infty\!\!\!d\omega f_1(\omega)\tanh\Big(\frac{\beta\omega}{2}\Big)\Big]
\frac{\tanh\left(\frac{\beta\omega_{M+1}}{2}\right)}{\omega_{M+1}}\Omega.
\end{align}
This expression easily demonstrates that the main contribution to the generated coherence comes from 
the domain $\omega\gg 2/\beta$, where $\tanh(\beta\omega/2)\approx 1$. Therefore one can use the simplified formula for the coherence 
\begin{align}
\langle\sigma_{M+1}^x\rangle\approx 
 -4f_2 \Big[\int_{2/\beta}^\infty d\omega f_1(\omega)\Big]
\frac{\tanh\left(\frac{\beta\omega_{M+1}}{2}\right)}{\omega_{M+1}}\Omega.
\end{align}
Advantageously, even a broad distribution of coupling parameters $f_1(\omega)$ in the frequency domain
can be sufficient to induce nearly $M$
times higher coherence if the function $f_1(\omega)$ is localized in the region well above 
$2/\beta$. 

Note that the calculation of the average value of $\sigma_{M+1}^z$ is more complicated, 
since it needs to take into account the denominator terms in the reduced density operator 
$\rho_S$ from Eq.~(\ref{eq:reduced_density_operator}). 
Such a derivation is presented in Appendix~\ref{app:sigma_z}.   

\section{Coherence multiplexing to two spins ($M+2$ case)}
\label{sec:coherence_multiplexing}

\begin{figure}
	\centering
	\includegraphics[width=\linewidth]{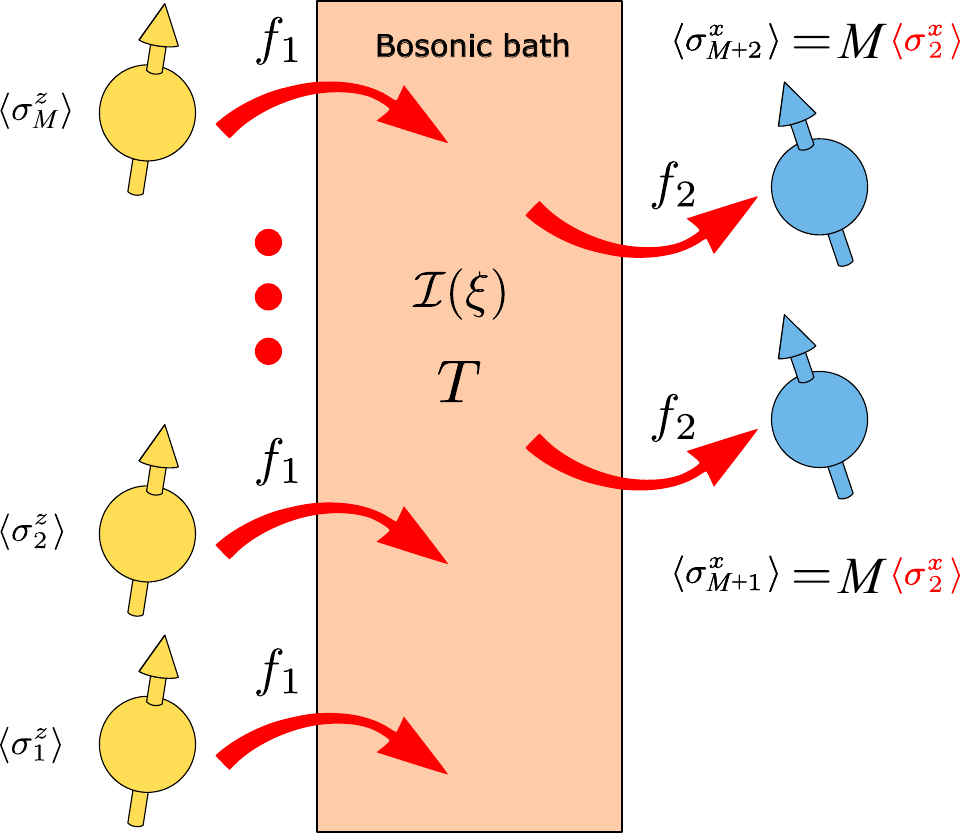}
	\caption{\label{fig:fig_2} Synthesizing and multiplexing of autonomous coherence.
	         The first group of $M$ driving spins  
	         polarizes the bosonic bath via the $\sum_{j=1}^M f_1\sigma_j^z$ term. The polarized bosonic bath
					 synthetically generates 
					 the coherence in the second group of the output spins via the $f_2\sigma_{M+p}^x$ terms, 
					 where index $p=1,2,\dots$
					 marks the target spins which get the coherence from the bath. For the coherence multiplexing, 
					 the figure represents the case of two output
					 spins. The generated coherences of these spins
					 $\langle\sigma_{M+1}^x\rangle=\langle\sigma_{M+2}^x\rangle$ are
					 $M$ times larger than the coherence $\langle\sigma_2^x\rangle$ (colored in red) 
					 generated in the direct two-spins scheme, see Fig.~\ref{fig:fig_1}(b)  
					 and Eq.~(\ref{eq:second_spin_coherence}).}
\end{figure}

Now we are in position to analyse another problem concerning detection of autonomous coherences, when more spins can be used as output systems. If they all independently receive the same coherence from a single bath, we can observe it more easily without need to repeat the experiment in time. For analysis of coherence multiplexing, we 
extend the previous study to the case of $(M+2)$ spins coupled to the bosonic bath, see Fig.~\ref{fig:fig_2}. 
The Hamiltonians of the spin system and spin-boson coupling term are    
\begin{equation}
H_S=\sum_{j=1}^M\frac{\omega_j}{2}\sigma_j^z+\sum_{p=1}^2\frac{\omega_{M+p}}{2}\sigma_{M+p}^z
%+\frac{\omega_{M+2}}{2}\sigma_{M+2}^z,
\end{equation} 
and  
\begin{equation}
H_{SB}=\Big[
\sum_{j=1}^M\boldsymbol{\sigma}_j\cdot\boldsymbol{n}_j+\sum_{p=1}^2\boldsymbol{\sigma}_{M+p}\cdot\boldsymbol{n}_{M+p}
%+\boldsymbol{\sigma}_{M+2}\boldsymbol{n}_{M+2}
\Big]\sum_k \lambda_k(b_k^\dag+b_k),
\end{equation} 
respectively. We again consider the special case $\boldsymbol{n}_j=(0,0,f_1)$, with $j=1,2,\dots M$, 
$\boldsymbol{n}_{M+1}=\boldsymbol{n}_{M+2}=(f_2,0,0)$. 
Repeating the averaging procedure over the bosonic degrees of freedom, see Appendix~\ref{app:derivation},
we obtain the following spin-spin correlation term
\begin{widetext}
\begin{align}
F(\tau)F(\tau')%=&\Big[f_1\sum_{j=1}^M\sigma_j^z+f_2\sigma_{M+1}^x(\tau)+f_2\sigma_{M+2}^x(\tau)\Big]
%\Big[f_1\sum_{j=1}^M\sigma_j^z+f_2\sigma_{M+1}^x(\tau')+f_2\sigma_{M+2}^x(\tau')\Big]=\nonumber \\ 
%=&f_1^2\sum_{j,l=1}^M\sigma_j^z\sigma_l^z+f_1f_2\sum_{j=1}^M [\sigma_{M+1}^x(\tau')+
%\sigma_{M+1}^x(\tau)+\sigma_{M+2}^x(\tau')+
%\sigma_{M+2}^x(\tau)]\sigma_j^z+ \nonumber \\+&f_2^2[\sigma_{M+1}^x(\tau)+\sigma_{M+2}^x(\tau)][\sigma_{M+1}^x(\tau')+\sigma_{M+2}^x(\tau')]=\nonumber 
%\\
=&f_1f_2\sum_{j=1}^M \sigma_j^z \sum_{p=1}^2 [\cosh(\omega_{M+p}\tau')+\cosh(\omega_{M+p}\tau)]\sigma_{M+p}^x+
\nonumber \\+&
if_1f_2\sum_{j=1}^M \sigma_j^z \sum_{p=1}^2 
%\nonumber \\&\qquad\qquad\qquad\quad+
[\sinh(\omega_{M+p}\tau')+\sinh(\omega_{M+p}\tau)]\sigma_{M+p}^y+\nonumber \\+&
f_1^2\sum_{j,l=1}^M\sigma_j^z\sigma_l^z+f_2^2\sum_{p=1}^2\Big[\cosh(\omega_{M+p}(\tau-\tau'))+
\sinh(\omega_{M+p}(\tau-\tau'))
\sigma_{M+p}^z\Big] +\nonumber \\+&
f_2^2\sigma_{M+1}^x\sigma_{M+2}^x [\cosh(\omega_{M+1}\tau)\cosh(\omega_{M+2}\tau')+
\cosh(\omega_{M+2}\tau)\cosh(\omega_{M+1}\tau')]-\nonumber \\-&
f_2^2\sigma_{M+1}^y\sigma_{M+2}^y [\sinh(\omega_{M+1}\tau)\sinh(\omega_{M+2}\tau')+
\sinh(\omega_{M+2}\tau)\sinh(\omega_{M+1}\tau')]+\nonumber \\+&
if_2^2\sigma_{M+1}^x\sigma_{M+2}^y [\cosh(\omega_{M+1}\tau)\sinh(\omega_{M+2}\tau')+
\sinh(\omega_{M+2}\tau)\cosh(\omega_{M+1}\tau')]+\nonumber \\+&
if_2^2\sigma_{M+1}^y\sigma_{M+2}^x [\sinh(\omega_{M+1}\tau)\cosh(\omega_{M+2}\tau')+
\cosh(\omega_{M+2}\tau)\sinh(\omega_{M+1}\tau')].
\end{align}
\end{widetext}
Consequently,
\begin{align}
\langle F(\tau)&F(\tau')\sigma_{M+p}^x\rangle_S=
-f_1f_2\sum_{j=1}^M \tanh\Big(\frac{\beta\omega_j}{2}\Big) \times \nonumber 
\\ \times&\Big\{[\cosh(\omega_{M+p}\tau')+\cosh(\omega_{M+p}\tau)]-%\nonumber \\ -&
[\sinh(\omega_{M+p}\tau')+\sinh(\omega_{M+p}\tau)]\tanh\Big(\frac{\beta \omega_{M+p}}{2}\Big)\Big\}.
\end{align}
After the substituting this result into the expression for the coherence of the $(M+p)$th spin one can observe
that it has the same structure as the coherence, obtained in Sec.~\ref{sec:synthesizing_coherence} for the $(M+1)$th spin. Therefore, the level of the coherence of the $(M+1)$th spin doesn't decrease 
the level of the coherence of $(M+2)$th spin and vice versa. However, the pair of output spins can be still correlated which will reduce their application as independent resources. In order to check the correlation between these spins we  calculate the correlation parameter
\begin{equation}
\label{eq:sigma_definition}
\sigma\equiv\sqrt{\langle\sigma_{M+1}^x\sigma_{M+2}^x\rangle-\langle\sigma_{M+1}^x\rangle\langle\sigma_{M+2}^x\rangle}.
\end{equation}
In the leading order in coupling parameters $f_1,f_2$ we get the following expression 
(see intermediate results of the calculation in Appendix~\ref{app:intermediate}) 
\begin{equation}
\label{eq:sigma_correlation}
\sigma^2\approx 4f_2^2\int_0^\infty d\xi\,\frac{\mathcal{I}(\xi)}{\xi} \mathcal{G}(\xi,\omega_{M+1},\omega_{M+2}), 
\end{equation}
where
\begin{align}
\label{eq:G}
\mathcal{G}(\xi,x,y)=&
\frac{xy\xi\tanh\left(\frac{\beta x}{2}\right)\tanh\left(\frac{\beta y}{2}\right)}
{\left(\xi^2-x^2\right)\left(\xi^2-y^2\right)}\coth\Big(\frac{\beta\xi}{2}\Big)
 + %\nonumber \\+&
\frac{xy\xi^2}{(x^2-y^2)}\Big[\frac{\tanh\left(\frac{\beta x}{2}\right)}{y(\xi^2-y^2)}-
 \frac{\tanh\left(\frac{\beta y}{2}\right)}
{x(\xi^2-x^2)}\Big].
\end{align}
One can observe that the integrand $\mathcal{G}(\xi,x,y)$ is a positive regular function on $
\xi\in[0,\infty)$ domain. It is a symmetric function of the parameters $x,y$ and its values 
belong to the domain $\mathcal{G}(\xi,x,y)\in[\mathcal{G}_{min}(x,y), 
\mathcal{G}_{max}(x,y)]$ with 
\begin{align}
\mathcal{G}_{min}(x,y)=&\frac{2 \tanh \left(\frac{\beta x}{2}\right) 
\tanh\left(\frac{\beta y}{2}\right)}{\beta xy},\\
\mathcal{G}_{max}(x,y)=&\frac{x\tanh\left(\frac{\beta x}{2}\right)
-y\tanh\left(\frac{\beta y}{2}\right)}{x^2-y^2}.
\end{align}
Therefore we have the following bands on the correlation value 
\begin{equation}
 \mathcal{G}_{min}(\omega_{M+1},\omega_{M+2})\leq\frac{\sigma^2}{4f_2^2\Omega}\leq 
 \mathcal{G}_{max}(\omega_{M+1},\omega_{M+2}).
\end{equation}
In the high temperature limit %both values have the same limit 
$\mathcal{G}_{min}(\omega_{M+1},\omega_{M+2}),
\mathcal{G}_{max}(\omega_{M+1},\omega_{M+2})\rightarrow \beta/2$. 
Therefore in this case we have the following answer 
\begin{equation}
\sigma^2=2\beta f_2^2\Omega.
\end{equation}
Note that the result has the universal character, i.e., it doesn't depend on 
the parameters of the spin systems $\omega_{M+1}, \omega_{M+2}$.  
In the low-temperature limit $\beta \omega_{M+1},\beta\omega_{M+2}\gg 1$ the $\sigma^2$ has a non-vanishing value 
\begin{equation}
\sigma^2=\frac{4f_2^2}{\omega_{M+1}+\omega_{M+2}}\int_0^\infty d\xi\,\mathcal{I}(\xi)
\frac{\xi+\omega_{M+1}+\omega_{M+2}}{(\xi+\omega_{M+1})(\xi+\omega_{M+2})}.
\end{equation}
Note that the non-zero correlation between both spins appears in the leading order of the perturbation theory and 
it can be interpreted as the reaction of the second spin to the first one (and vice versa) using the states of the  bosonic bath as intermediate virtual states. Indeed, the leading term of the correlation value contains only the  coupling parameter $f_2$, but not the coupling parameters $f_1$, so the non-zero correlation value will be also at 
$f_1=0$.  

In order to understand the effectiveness of such a scheme, we estimate the signal-to-noise ratio 
$|\langle\sigma_{M+1}^x\rangle|/\sigma$. 
We consider the case of the input and output spins with the same frequencies, 
namely, $\omega_j=\omega_1$ for $j=1,2,\dots M$ and $\omega_{M+1}=\omega_{M+2}=\omega$, as a typical example    
\begin{align}
\frac{|\langle\sigma_{M+1}^x\rangle|}{\sigma}=\frac{2f_1\Omega M\tanh\Big(\frac{\beta\omega_1}{2}\Big)
\tanh\Big(\frac{\beta\omega}{2}\Big)}{\omega \sqrt{\int_0^\infty d\xi \frac{\mathcal{I}(\xi)}{\xi}\mathcal{G}(\xi,\omega,\omega)}}.
\end{align}
Using the special case of the power-law (generalized Ohmic) spectral density function (\ref{eq:spectral}) we obtain the following expression
\begin{align}
\frac{\langle\sigma_{M+1}^x\rangle}{\sigma}=\eta
\frac{\Big(\frac{\omega_c}{\omega_1}\Big)^{\frac{s+1}{2}} 
\Gamma(s)\tanh\Big(\frac{b}{2}\Big)\tanh\Big(\frac{bw}{2}\Big)}
{\sqrt{w}\sqrt{\int_0^\infty dx x^s e^{-x\frac{\omega_1}{\omega_c}}g(x)}}. 
\end{align}
Here we have introduced the dimensionless parameter $\eta=4f_1M\sqrt{\lambda}$,
which doesn't depend on the input and output spin frequencies $\omega_1,\omega$ and 
inverse temperature $\beta$,
\begin{align}
g(x)=&\frac{\tanh^2\left(\frac{bw}{2}\right)\left[bwx\left(w^2-x^2\right)+4w^3\coth\left(\frac{bx}{2}\right)\right]}
{\left(w^2-x^2\right)^2}
+\nonumber \\+&
\frac{bwx\left(x^2-w^2\right)+\left(2x^3-6w^2x\right)\tanh\left(\frac{bw}{2}\right)}{\left(w^2-x^2\right)^2},
\end{align}
and short notations $w=\omega/\omega_1$, $b=\beta\omega_1$ for brevity. 

\begin{figure*}[t]
	\centering
	\includegraphics[width=\linewidth]{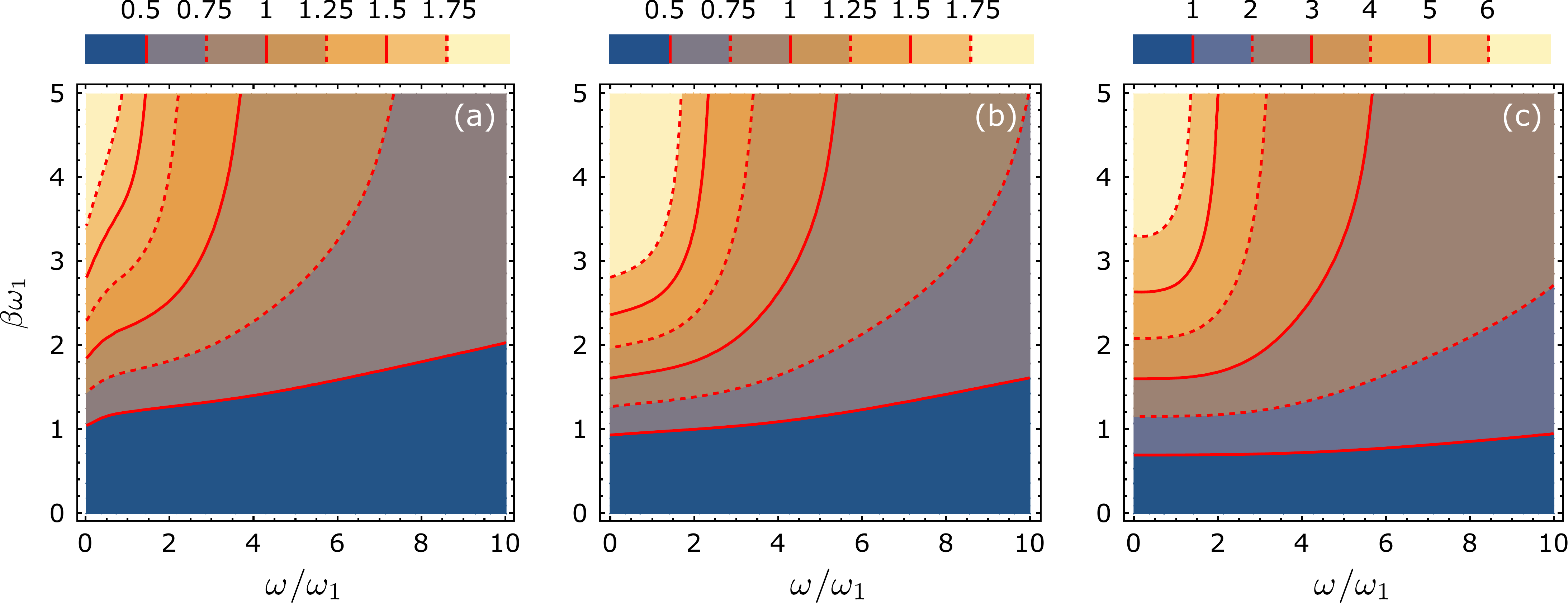} 
\caption{Density plot of normalized signal-to-noise ratio $\langle\sigma_{M+1}^x\rangle/\eta\sigma$ 
        for (a) sub-Ohmic with $s=0.5$, (b) Ohmic with $s=1$, and (c) super-Ohmic with $s=2$ cases
				in the domain $\omega/\omega_1 \in[0,10]$ and $\beta\omega_1 \in[0,5]$. }
\label{fig:fig_signal}				
\end{figure*}

Using the obtained formula we plot  in Fig.~\ref{fig:fig_signal} the normalized signal-to-noise ratio 
$\langle\sigma_{M+1}^x\rangle/\eta\sigma$ as a function of the dimensionless parameters 
$\beta\omega_1$ and $\omega/\omega_1$ for different values of $s=0.5;1;2$.   
    
\section{Synthesizing coherence from two independent baths ($M+N+1$ case)}
\label{sec:two_baths}
Now, the last intriguing problem remains for our new method: do we really need a single bath that is coherently manipulated by $M$ driving spins? Cannot we use two different and independent baths coherently pushed by $M$ and $N$ spins if they both transfer coherence through these baths to the same output spin?

\begin{figure*}
	\centering
	\includegraphics[width=\linewidth]{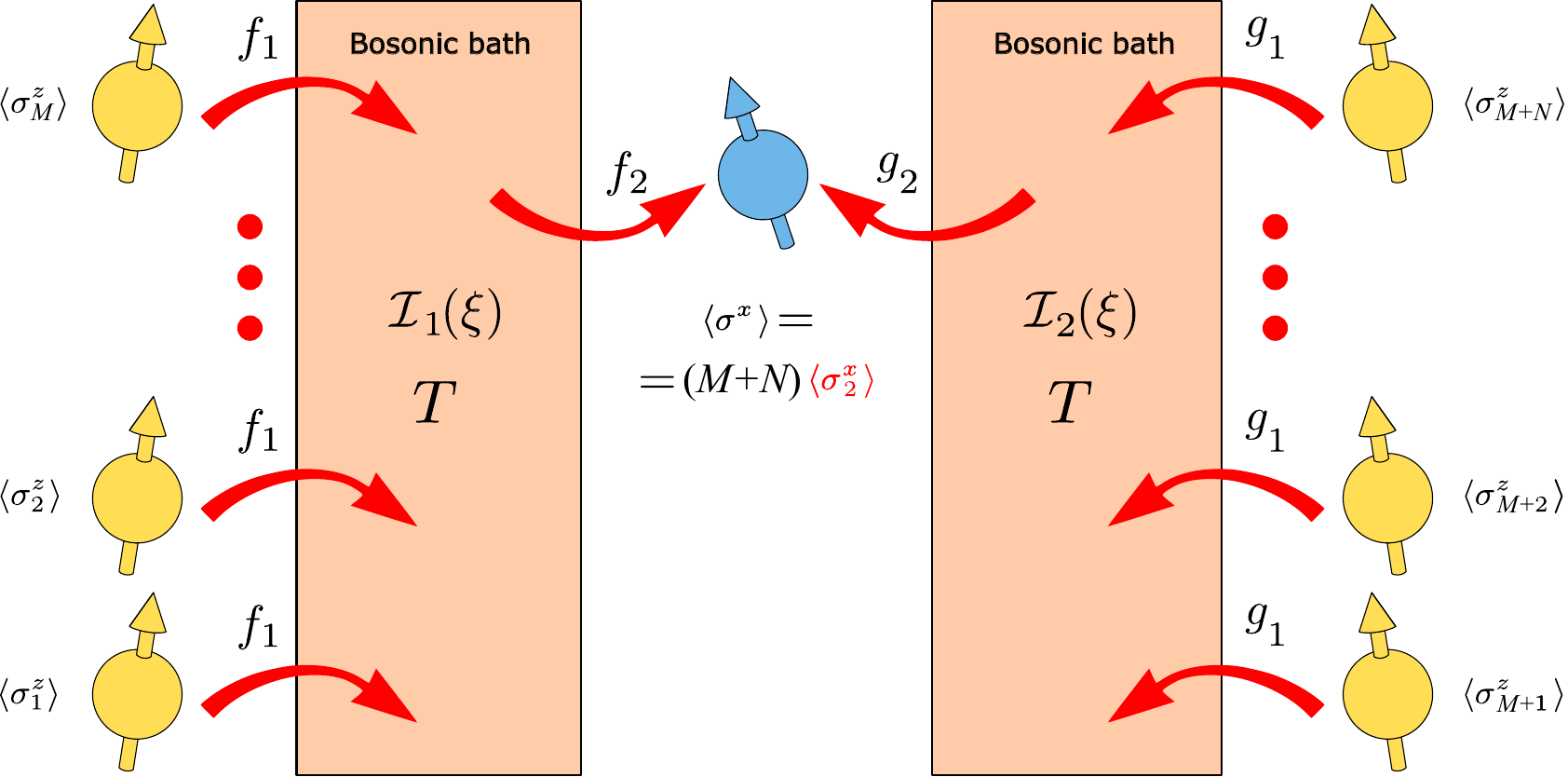}
	\caption{\label{fig:fig_3} Autonomous quantum coherence from several independent baths. 
	         The group of first $M$ driving spins polarizes the first bosonic bath 
					 via $\sum_{j=1}^M f_1\sigma_j^z$ term, 
					 the group of another $N$ driving spins polarizes the second bosonic bath via 
					 $\sum_{j=M+1}^{M+N} g_1\sigma_j^z$ term. The polarized baths simultaneously generate the coherence 
					 $\langle\sigma^x\rangle$ of the $(M+N+1)$th output spin via $f_2\sigma^x$ and $g_2\sigma^x$ terms.
					 The coherence of this spin in the leading order in the coupling parameters $f_1,f_2,g_1,g_2$ 
					 is $(M+N)$ times larger than the coherence $\langle\sigma_2^x\rangle$ 
					 (colored in red) generated in the direct two-spins scheme, see 
					 Fig.~\ref{fig:fig_1}(b) and Eq.~(\ref{eq:second_spin_coherence}).}
\end{figure*}

We consider an extended version of the previous system, see Fig.~\ref{fig:fig_3}. 
We introduce two reservoirs of bosons, 
with spectra $\Omega_k,\Upsilon_k>0$ and creation (annihilation) operators $b_k^\dag,d_k^\dag (b_k,d_k)$,
respectively.  
First (second) reservoir is coupled to $M$ ($N$) spins via $f_1\sigma_j^z$($g_1\sigma_j^z$) term.
Both reservoirs are coupled to the $(M+N+1)$th spin via $f_2\sigma^x$ and $g_2\sigma^x$ terms. 
The Hamiltonian of the system 
\begin{align}
\label{eq:full_hamiltonian}
H=H_S+H_B+H_{SB}
\end{align}
consists of the spin Hamiltonian
\begin{align}
H_S=&\sum_{j=1}^M\frac{\omega_1}{2}\sigma_j^z+\sum_{j=M+1}^{M+N} 
\frac{\omega_2}{2}\sigma_j^z+\frac{\omega}{2}\sigma^z,
\end{align}
the bosnic Hamiltonian
\begin{align}
H_B=&\sum_k\Omega_k b_k^\dag b_k + \sum_k \Upsilon_k d_k^\dag d_k,
\end{align}
and the spin-bath interaction term 
\begin{align}
H_{SB}=&f_1\sum_{j=1}^M\sigma_j^z \sum_k \lambda_k(b_k^\dag+b_k)+%\nonumber \\+&
g_1\sum_{j=M+1}^{M+N}\sigma_j^z \sum_k \kappa_k(d_k^\dag+d_k)+ \nonumber \\+&
\sigma^x \sum_{k} \Big[ f_2\lambda_k (b^\dag_k+b_k)+g_2\kappa_k (d^\dag_k+d_k)\Big].
\end{align}
The first line of the $H_{SB}$ describes the interaction of the $M$ input spins with the left bath, 
the second line describes the interaction of the additional $N$ input spins with the right bath, 
and finally the last term in the $H_{SB}$ describes the coupling of the remaining $(M+N+1)$th spin 
with both baths. The first and second baths are characterized by spectral density functions 
\begin{equation}
\label{eq:spectral_density_functions}
\mathcal{I}_1(\xi)=\sum_k \lambda_k^2\delta(\xi-\Omega_k),\quad 
\mathcal{I}_2(\xi)=\sum_k \kappa_k^2\delta(\xi-\Upsilon_k),
\end{equation} 
the reorganization energies 
\begin{align}
\Omega=\int_0^\infty d\xi\,\mathcal{I}_1(\xi)/\xi, \quad 
\Upsilon=\int_0^\infty d\xi\,\mathcal{I}_2(\xi)/\xi).
\end{align}
We are interested in the evaluation of the average 
\begin{align}
\langle\sigma^x\rangle=\frac{1}{Z}\text{Tr}\Big[e^{-\beta H}\sigma^x\Big],
\end{align}
where  $Z\equiv\text{Tr}[e^{-\beta H}]$. The coherence $\langle\sigma^x\rangle$ reads 
(details of the calculation are presented in the Appendix~\ref{app:sigma_x}) 
\begin{widetext}
\begin{align}
\label{eq:coherence_sigma}
\langle\sigma^x\rangle=&\frac{Z_B}{Z}
\sum_{m=0}^M\sum_{n=0}^N {M \choose m}{N \choose n}
\exp\Big\{-\beta\big[\frac{\omega_1}{2}(M-2m)+\frac{\omega_2}{2}(N-2n)\big]\Big\}
\times \nonumber \\ \times & 
\exp\Big\{\beta\big[\Omega f_1^2(M-2m)^2+\Upsilon g_1^2(N-2n)^2\big]\Big\}
\text{Tr}_S\Big[
\exp\Big\{-\beta\frac{R_{mn}}{2}\sigma^z\Big\}
\times \nonumber \\ \times &
T_\tau \exp\Big\{f_2^2\int_0^\infty d\xi\,\mathcal{I}_1(\xi) 
\int_0^\beta d\tau \int_0^\beta d\tau' G(\xi,\tau-\tau')
\Sigma^x(\tau)\Sigma^x(\tau')\Big\}
\times \nonumber \\ \times &
T_\tau \exp\Big\{g_2^2\int_0^\infty d\xi\,\mathcal{I}_2(\xi) 
\int_0^\beta d\tau \int_0^\beta d\tau' G(\xi,\tau-\tau')
\Sigma^x(\tau)\Sigma^x(\tau')\Big\}\times \nonumber \\ \times& 
(\cos\theta_{mn}\sigma^x-\sin\theta_{mn}\sigma^z)\Big].
\end{align}
\end{widetext}
Here we introduced the partition function of the bosonic system $Z_B$, and the 
notations $R_{mn}=(\omega^2+\omega^2_{mn})^{1/2}$, 
with $\omega_{mn}=4f_1f_2\Omega(M-2m)+4g_1g_2\Upsilon(N-2n)$,
$\sin\theta_{mn}=\omega_{mn}/R_{mn}$ and $\cos\theta_{mn}=\omega/R_{mn}$. 
Furthermore, $\Sigma^x(\tau)=\cosh(R_{mn}\tau)[\cos\theta_{mn}\sigma^x+
i\cos\theta_{mn}\sigma^y]-\sin\theta_{mn}\sigma^z$, and $G(\xi, \tau-\tau')$ is 
defined in Eq.~(\ref{eq:Green_function}).
%\begin{widetext}
%\begin{equation}
%\Sigma^x(\tau)=\cos\theta_{mn}[\cosh(R_{mn}\tau)\sigma^x+i\sinh(R_{mn}\tau)\sigma^y]-\sin\theta_{mn}\sigma^z.
%\end{equation}	
%\end{widetext}
%\begin{align}
%G(\xi,\tau-\tau')=
%\frac{e^{(\tau-\tau')\xi}}{e^{\beta\xi}-1}\theta(\tau-\tau')+
%\frac{e^{(\tau-\tau')\xi}}{1-e^{-\beta\xi}}\theta(\tau'-\tau),
%\end{align}
%is the bosonic Green function. 
Finally, the partition function $Z$ of the whole system is 
\begin{widetext}
\begin{align}
\label{eq:partition_function}
Z=&Z_B
\sum_{m=0}^M\sum_{n=0}^N {M \choose m}{N \choose n}
\exp\Big\{-\beta\big[\frac{\omega_1}{2}(M-2m)+\frac{\omega_2}{2}(N-2n)\big]\Big\}
\times \nonumber \\ \times &
\exp\Big\{\beta\big[\Omega f_1^2(M-2m)^2+\Upsilon g_1^2(N-2n)^2\big]\Big\}
\text{Tr}_S\Big[\exp\Big\{-\beta\frac{R_{mn}}{2}\sigma^z\Big\}
\times \nonumber \\ \times &
T_\tau \exp\Big\{f_2^2\int_0^\infty d\xi\,\mathcal{I}_1(\xi) \int_0^\beta d\tau \int_0^\beta d\tau' G(\xi,\tau-\tau')
\Sigma^x(\tau)\Sigma^x(\tau')\Big\}
\times \nonumber \\ \times &
T_\tau \exp\Big\{g_2^2\int_0^\infty d\xi\,\mathcal{I}_2(\xi) \int_0^\beta d\tau \int_0^\beta d\tau' G(\xi,\tau-\tau')
\Sigma^x(\tau)\Sigma^x(\tau')\Big\}\Big].
\end{align} 	
\end{widetext}
Note that the perturbative regime corresponds to the situation $\omega\gg [\omega_{mn}]_\text{max}=
4f_1f_2\Omega M+4g_1g_2\Upsilon N$. In this limit $R_{mn}\approx \omega$, 
$\sin\theta_{mn}\approx [4f_1f_2\Omega(M-2m)+4g_1g_2\Upsilon(N-2n)]/\omega$, 
$\cos\theta_{mn}\approx 1$. 

Let's check this formula for the previously considered case of $M+1$ spins, 
where $4f_1f_2\Omega M\ll\omega$, $g_2=0$.
For this case the expression for the statistical sum takes the form 
\begin{widetext}
\begin{align}
Z=&Z_B
\sum_{m=0}^M\sum_{n=0}^N {M \choose m}{N \choose n}
\exp\Big\{-\beta\big[\frac{\omega_1}{2}(M-2m)+\frac{\omega_2}{2}(N-2n)\big]\Big\}
\times \nonumber \\ \times &
\exp\Big\{\beta\big[\Omega f_1^2(M-2m)^2+\Upsilon g_1^2(N-2n)^2\big]\Big\}
\text{Tr}_S\Big[\exp\Big\{-\beta\frac{R_{m}}{2}\sigma^z\Big\}
\times \nonumber \\ \times &
T_\tau \exp\Big\{f_2^2\int_0^\infty d\xi\,\mathcal{I}_1(\xi) \int_0^\beta d\tau \int_0^\beta d\tau' G(\xi,\tau-\tau')
\Sigma^x(\tau)\Sigma^x(\tau')\Big\}\Big],
\end{align} 
\end{widetext}
where we introduced $R_m=\sqrt{\omega^2+[4f_1f_2\Omega(M-2m)]^2}$. We keep the leading term in $T_\tau$ exponent, i.e., 
replace it by $1$. Then we use the approximation $R_{m}\approx \omega$, 
and remove the small terms $\propto f_1^2\Omega,g_1^2\Upsilon$ 
in the expression for the exponent. Then we get 
\begin{widetext}
\begin{align}
Z\approx &Z_B
\sum_{m=0}^M\sum_{n=0}^N {M \choose m}{N \choose n}
\exp\Big\{-\beta\big[\frac{\omega_1}{2}(M-2m)+
\frac{\omega_2}{2}(N-2n)\big]\Big\}
\text{Tr}_S\Big[\exp\Big\{-\beta\frac{\omega}{2}\sigma^z\Big\}
\Big]
\nonumber \\ =&
Z_B\Big[2\cosh\Big(\frac{\beta\omega}{2}\Big)\Big]\Big[2\cosh\Big(\frac{\beta\omega_1}{2}\Big)\Big]^M
\Big[2\cosh\Big(\frac{\beta\omega_2}{2}\Big)\Big]^N,
\end{align} 		
\end{widetext}
which is nothing but the statistical sum of the non-interacting spin and boson subsystems. 
Using the same approximation and $\sin\theta_m=4f_1f_2\Omega(M-2m)/R_m\approx 4f_1f_2\Omega(M-2m)/\omega$
we obtain the following expression for the coherence in the system 
\begin{widetext}
\begin{align}
\langle\sigma^x\rangle\approx&\frac{Z_B}{Z}
\sum_{m=0}^M\sum_{n=0}^N {M \choose m}{N \choose n}
\exp\Big\{-\beta\big[\frac{\omega_1}{2}(M-2m)+
\frac{\omega_2}{2}(N-2n)\big]\Big\}\times \nonumber \\ \times& 
\text{Tr}_S\Big[
\exp\Big\{-\beta\frac{\omega}{2}\sigma^z\Big\}
(-\sin\theta_m\sigma^z)\Big]= 
\frac{Z_B}{Z}\frac{4f_1f_2\Omega}{\omega}
\Big[2\sinh\Big(\frac{\beta\omega}{2}\Big)\Big]\Big(-\frac{2}{\beta}\Big) 
\times \nonumber \\ \times&
\frac{\partial}{\partial\omega_1}\sum_{m=0}^M\sum_{n=0}^N {M \choose m}{N \choose n}
\exp\Big\{-\beta\big[\frac{\omega_1}{2}(M-2m)+
\frac{\omega_2}{2}(N-2n)\big]\Big\}%=\nonumber \\ =&-
%\frac{Z_{B1}Z_{B2}}{Z}\frac{8f_1f_2\Omega}{\beta\omega}
%\Big[2\sinh\Big(\frac{\beta\omega}{2}\Big)\Big]
%\Big[2\cosh\Big(\frac{\beta\omega_2}{2}\Big)\Big]^N 
%\frac{\partial}{\partial\omega_1}\Big[2\cosh\Big(\frac{\beta\omega_1}{2}\Big)\Big]^M
=\nonumber \\ =& 
-\frac{4f_1f_2\Omega}{\omega}\tanh\Big(\frac{\beta\omega}{2}\Big)M\tanh\Big(\frac{\beta\omega_1}{2}\Big),
\end{align} 
\end{widetext}		 
which coincides with the previously obtained result (\ref{eq:multispin_result}). 
Considering the general case with $g_2\neq 0$ in the same approximation we get 
\begin{widetext}
\begin{align}
\langle\sigma^x\rangle \approx -4\frac{\tanh\Big(\frac{\beta\omega}{2}\Big)}{\omega}
\Big[f_1f_2\Omega M\tanh\Big(\frac{\beta\omega_1}{2}\Big)+g_1g_2\Upsilon N\tanh\Big(\frac{\beta\omega_2}{2}\Big)\Big].
\end{align} 
\end{widetext}
Therefore, the impact of several baths on the coherence of the $\sigma$'s spin has an additive character.
It means that independent baths can constructively join to generate coherence of the output spins.

Note the several advantages of the results (\ref{eq:coherence_sigma}), (\ref{eq:partition_function}),
and (\ref{eq:reduced}): i) it can be easily generalized for the arbitrary number of baths; 
ii) it is non-perturbative in the coupling parameters $f_,f_2,g_1,g_2$; 
iii) it allows to obtain the answers beyond the weak coupling limit 
$4f_1f_2\Omega M +4g_1g_2\Upsilon N\ll \omega$, 
$2f_1^2\Omega M\ll \omega_1$, $2g_1g_2\Upsilon N\ll \omega_2$. However, 
the evaluation of the average of the coherence beyond the weak coupling limit 
requires either more sophisticated analytical methods or direct numerical calculations.  		

\section{Generalization for the oscillator coherences}
\label{sec:generalization}

To extend the experimental possibilities we finally 
replace the output spin system of the previous case by an oscillator. 
The simplest model Hamiltonian of the corresponding system analogously to Sec.~\ref{sec:multispin}
is $H=H_A+H_S+H_B+H_{int}$, where
\begin{align}
H_A=E\Big(a^\dag a+\frac12\Big),
\end{align}
is the oscillator Hamiltonian, 
\begin{align}
H_B=\sum_k \Omega_k b_k^\dag b_k,
\end{align}
the Hamiltoinan of the bosonic excitations of the bath, 
\begin{align}
H_S=\frac{\omega}{2}\sigma^z,
\end{align}
is the spin Hamiltonian and finally the term
\begin{align}
H_{int}=\Big[f\sigma^z+g(a^\dag+a)\Big]\sum_k \lambda_k(b_k^\dag+b_k),
\end{align}
describes an interaction of the spin and the oscillator with the bosonic bath. 

Applying the general formula for this case we obtain
\begin{widetext}
\begin{equation}
\label{eq:rho_a}
\rho_A=\frac{e^{-\beta H_{A}}}{Z_A}
\frac{\Big\langle T_\tau \exp\Big\{\int_0^\infty d\xi\,\mathcal{I}(\xi) \int_0^\beta d\tau \int_0^\beta d\tau' G(\xi,\tau-\tau')
F(\tau)F(\tau')\Big\}\Big\rangle_S}
{\Big\langle\Big\langle T_\tau \exp\Big\{\int_0^\infty d\xi\,\mathcal{I}(\xi) \int_0^\beta d\tau \int_0^\beta d\tau' G(\xi,\tau-\tau')
F(\tau)F(\tau')\Big\}\Big\rangle_S\Big\rangle_A},  
\end{equation}
\end{widetext}
where we have introduced 
\begin{align}
F(\tau)=f\sigma^z+g(a^\dag e^{\tau E}+a e^{-\tau E})
\end{align}
The reduced density matrix of the oscillator can be calculated  as a power series in the parameters $f,g$. 
We calculate the leading order of the $T_\tau$-exponent in the numerator of Eq.~(\ref{eq:rho_a})
\begin{widetext}
\begin{align}
\Big\langle T_\tau \exp\{\dots\}\Big\rangle_S%\approx& 1+ 
%\int_0^\infty d\xi\,\mathcal{I}(\xi) \int_0^\beta d\tau \int_0^\tau d\tau' \phi(\xi,\tau-\tau')
%\Big\{-fg\tanh\Big(\frac{\beta\omega}{2}\Big)\Big[a^\dag \Big(e^{\tau E}+e^{\tau' E}\Big)+a \Big(e^{-\tau E}+e^{-\tau' E}\big)\Big]+
%  \nonumber \\
%&+
%g^2 \Big[(a^\dag )^2 e^{(\tau+\tau')E}+a^2 e^{-(\tau+\tau')E} +
%a^\dag a e^{(\tau-\tau')E}+aa^\dag e^{-(\tau-\tau')E}\Big] + f^2 \Big\} =\nonumber \\=&
\approx& 1+ \int_0^\infty d\xi\,\mathcal{I}(\xi) \Big\{\frac{f^2\beta}{\xi}-\frac{2fg}{E\xi}\tanh\Big(\frac{\beta\omega}{2}\Big)
\Big[a^\dag(e^{\beta E}-1) +a (1-e^{-\beta E})\Big]+ \nonumber \\+&
g^2 \Big[\frac{E\coth\left(\frac{\beta\xi}{2}\right)-
\xi\coth\left(\frac{\beta E}{2}\right)}{2E\left(E^2-\xi^2\right)}\Big\{(a^\dag )^2\left(e^{\beta E }-1\right)^2  +a^2 \left(1-e^{-\beta E}\right)^2\Big\} +\nonumber \\+&
\Big(\frac{\beta(\xi-E)+e^{\beta(E-\xi)}-1}{\left(1-e^{-\beta\xi}\right)(E-\xi)^2}
+\frac{-\beta(\xi+E)+e^{\beta(\xi+E)}-1}{\left(e^{\beta\xi}-1\right) (\xi+E)^2}\Big)a^\dag a + \nonumber \\+&
\Big(\frac{\beta(E-\xi)+e^{\beta(\xi-E)}-1}{\left(e^{\beta\xi}-1\right) (E-\xi)^2}+\frac{\beta(\xi +E)+e^{-\beta(\xi+E)}-1}{\left(1-e^{-\beta\xi}\right) (\xi+E)^2}\Big)aa^\dag \Big]
\Big\}.
\end{align}
\end{widetext}
The denominator of the reduced density matrix then reads
\begin{widetext}
\begin{align}
\Big\langle T_\tau \exp\{\dots\}\Big\rangle_{A,S}\approx 1+ \int_0^\infty d\xi\,\mathcal{I}(\xi) \Big\{\frac{f^2\beta}{\xi}+g^2 \frac{\beta\left[E\coth\left(\frac{\beta\xi}{2}\right)-\xi\coth\left(\frac{\beta E}{2}\right)\right]}{E^2-\xi ^2} 
\Big\}
\end{align}
\end{widetext}
Let us use the obtained result for $\rho_A$ and calculate the average values for the dimensionless
coordinate operator $x\equiv(a^\dag+a)/\sqrt{2}$
%and its square as an example. To do it we use the following expression for the coordinate operator 
%\begin{align}
%x\def\sqrt{1}{2}}(a^\dag+a). %\quad p=i\sqrt{\frac{\hbar mw}{2}}(a^\dag-a),
%\end{align}
\begin{equation}
\langle x\rangle
\approx -\frac{4fg}{\sqrt{2}E}\Omega\tanh\Big(\frac{\beta\omega}{2}\Big).
%\Big\langle T_\tau \exp\{\dots\}\Big\rangle^{-1}_{A,S}.
\end{equation}
Note that the non-zero value of $x$ operator exists if both values $g$ and $f$ are non-zero. 
This result can be understood in terms of a two-step model. First, the spin polarizes the bosonic bath 
via $f\sigma^z$ term and produces the non-zero value $\langle b_k+b_k^\dag\rangle \propto f$ 
in the leading order of the perturbation theory. Then the polarized bath generates the non-zero coordinate
shift $\langle a^\dag+a\rangle\propto g \langle b_k+b_k^\dag\rangle \propto gf$ via the $ g (a^\dag +a)$ term.
The logic of the method advantageously follows the previous case of the output spin.
The square of the coordinate operator takes the form 
\begin{widetext} 
\begin{align}
\langle x^2\rangle\approx &
%\frac12\Big\{\coth\Big(\frac{\beta E}{2}\Big)+f^2\beta \coth\Big(\frac{\beta E}{2}\Big)\Omega+
%g^2\int_0^\infty d\xi\,\mathcal{I}(\xi)\frac{4E}{\left(E^2-\xi^2\right)^2}
%\left[E\coth\left(\frac{\beta\xi}{2}\right)-\xi \coth\left(\frac{\beta E}{2}\right)\right]+\nonumber \\
%+&
%g^2\int_0^\infty d\xi\frac{\mathcal{I}(\xi)}{\left(E^2-\xi^2\right)}
%\left[\beta E\coth\left(\frac{\beta\xi}{2}\right)\coth\left(\frac{\beta E}{2}\right)+ \beta\xi-
%2\beta\xi\coth^2\left(\frac{\beta E}{2}\right)-2\frac{\xi}{E}\coth\left(\frac{\beta E}{2}\right)\right]\Big\}
%\times \nonumber \\
%&\times\Big\langle T_\tau \exp\{\dots\}\Big\rangle^{-1}_{A,S}\approx\nonumber \\ \approx&
\frac{1}{2}\coth\Big(\frac{\beta E}{2}\Big)+
2E g^2\int_0^\infty\!\! d\xi\mathcal{I}(\xi)\frac{E\coth\left(\frac{\beta\xi}{2}\right)-
\xi \coth\left(\frac{\beta E}{2}\right)}{\left(E^2-\xi^2\right)^2}-\nonumber \\ -&
\frac{g^2}{2E}\frac{\beta E+\sinh(\beta E)}{\sinh^2(\beta E/2)}
\int_0^\infty\!\! d\xi\frac{\mathcal{I}(\xi) \xi}{\left(E^2-\xi^2\right)}.
\end{align}
\end{widetext}
As a result for the variance $\sigma^2\equiv \langle(x-\langle x\rangle)^2\rangle$
we have $\sigma^2\approx \langle x^2\rangle$ in the leading order.
%\begin{widetext}
%\begin{align}
%\sigma^2\approx \Big[\coth\Big(\frac{\beta E}{2}\Big)+
%4E g^2\int_0^\infty\!\! d\xi\mathcal{I}(\xi)\frac{E\coth\left(\frac{\beta\xi}{2}\right)-
%\xi \coth\left(\frac{\beta E}{2}\right)}{\left(E^2-\xi^2\right)^2}
%-\frac{g^2}{E}\frac{\beta E+\sinh(\beta E)}{\sinh^2(\beta E/2)}
%\int_0^\infty\!\! d\xi\frac{\mathcal{I}(\xi) \xi}{\left(E^2-\xi^2\right)}\Big]
%\end{align}
%\end{widetext}
Note that the spin subsystem doesn't have an impact onto $\sigma^2$ in the leading order 
of the perturbation theory in coupling parameters $f,g$. 

Using the latter results one can estimate the signal-to-noise ratio $|\langle x\rangle|/\sigma$, 
which, in particular, has a simple form in the low-temperature limit $\beta(=1/T)\rightarrow \infty$
\begin{align}
\left.\frac{|\langle x\rangle|}{\sigma}\right|_{T\rightarrow 0}\approx 4fg\frac{\Omega}{E}. 
\end{align}
Hence, the signal-to-noise ratio in such a system can be increased by increaing the ratio of the 
reorganization energy of the bosonic bath $\Omega$ to the oscillator's energy $E$.

\section{Summary}
\label{sec:summary}

To conclude, using a new, more efficient method then in previous case, we proved that autonomous coherences could be synthesised from many independent driving spins through separate low-temperature baths and multiplxed to many output spins. It is a crucial step to allow their experiment at investigation for a much more diverse class of experimental platforms and final verification of such intriguing phenomena. The driving coupling is present as pure dephasing, for example, in quantum dots. Our analysis shows that another output spin system embedded in the same environment can exhibit such autonomous coherence in principle, although it cannot generate it. Moreover, more driving spins will be advantageous, and the same bath can supply many spins with the same autonomous coherence. Additionally, it is not required that driving spins have to generate coherence through the same bath; many baths coupled to the output spin will be equally good. 

Our approach qualitatively overcomes the first theoretical proposal suggesting the existence of these new autonomous spin coherences in Refs.~\cite{Guarnieri2018,Purkayastha2020,Slobodeniuk2022}.  The groundbreaking idea there is limited by a double role of spin in this method and, therefore, unavoidable disruptive back-action effects. We removed that limitation and opened a road towards further much broader investigations. This new approach and extension are critical to observing autonomous quantum coherences experimentally and exploiting them in diverse quantum technology applications.

\section{Acknowledgments}

R.F. acknowledges the grant of 22-27431S of the Czech Science Foundation. 
A.S. and T.N. acknowledges the grant by the Czech Science Foundation (project GA\v{C}R 23-06369S).

\bibliographystyle{plain}

\newpage

%\onecolumn\newpage
\appendix

\begin{widetext}

\section{Evaluation of the $T_\tau$-exponent}
\label{app:derivation}

To evaluate the reduced density operator in Eq.~(\ref{eq:reduced_density_operator}) we need to perform 
the averaging over bosonic degrees of freedom of the $T_\tau$ ordered exponent (\ref{eq:t_ordered}). 
To do it we present the spin-boson interaction term (\ref{eq:spin_boson}) as
\begin{equation}
H_{SB}\equiv\Big(\sum_{j=1}^{M+1}\boldsymbol{\sigma}_j\cdot\boldsymbol{n}_j\Big)\sum_k \lambda_k(b_k^\dag+b_k)
\equiv F\sum_k \lambda_k(b_k^\dag+b_k)\equiv\sum_k (f_kb_k^\dag+f_kb_k),
\end{equation}
and use the commutativity of the operators under the $T_\tau$ ordering operation
we present Eq.~(\ref{eq:t_ordered}) in the form 
\begin{align}
T_\tau\Big\{e^{-\int_0^\beta d\tau \widetilde{H}_{SB}(\tau)}\Big\}=&
T_\tau \Big\{e^{-\int_0^\beta d\tau \sum_k [f_k(\tau)b^\dag_k(\tau)+f_k(\tau)b_k(\tau)]}\Big\}=
\nonumber \\=&
T_\tau\prod_k \Big\{e^{-\int_0^\beta d\tau f_k(\tau)b^\dag_k(\tau)-\int_0^\beta d\tau' f_k(\tau')b_k(\tau')}\Big\}=\nonumber \\=&
T_\tau \prod_k \Big[\Big\{e^{-\int_0^\beta d\tau f_k(\tau)b^\dag_k(\tau)}\Big\}
\Big\{e^{-\int_0^\beta d\tau' f_k(\tau')b_k(\tau')}\Big\}\Big].
\end{align}

Then we the averaging over bosonic degrees of freedom of Eq.~(\ref{eq:t_ordered}) takes the form       
\begin{align}
\label{eq:expression}
\Big\langle T_\tau&\Big\{e^{-\int_0^\beta d\tau \widetilde{H}_{SB}(\tau)}\Big\}\Big\rangle_B=
\prod_k\sum_{n,m=0}^\infty \frac{(-1)^{n+m}}{n!m!}
\int_0^\beta \dots\int_0^\beta d\tau_1 \dots d\tau_n  
\int_0^\beta\dots\int_0^\beta d\tau'_l \dots d\tau'_m 
%\int_0^\beta d\tau_1 \dots \int_0^\beta d\tau_n\int_0^\beta d\tau'_1 \int_0^\beta \dots \int_0^\beta d\tau'_m 
\times \nonumber \\ 
\times&
T_\tau \Big\{f_k(\tau_1)\dots f_k(\tau_n)f_k(\tau'_1)\dots f_k(\tau'_m)\Big\}
\Big\langle T_\tau\Big\{b_k^\dag(\tau_1)\dots b_k^\dag(\tau_n)b_k(\tau'_1)\dots b_k(\tau'_m)\Big\}\Big\rangle_B=
\nonumber  \\
=&\prod_k\sum_{n=0}^\infty\frac{1}{(n!)^2}
%\int_0^\beta d\tau_1 \dots \int_0^\beta d\tau_n\int_0^\beta d\tau'_1 \dots \int_0^\beta d\tau'_n 
\prod_{j=1}^n \int_0^\beta\int_0^\beta d\tau_jd\tau'_j\, 
T_\tau \Big\{f_k(\tau_1)\dots f_k(\tau_n)f_k(\tau'_1)\dots f_k(\tau'_n)\Big\}
\times \nonumber \\ 
\times&
\sum_P \Big\langle T_\tau b_k^\dag(\tau_1)b_k(\tau'_{j_1})\Big\rangle_B \dots 
\Big\langle T_\tau b_k^\dag(\tau_n)b_k(\tau'_{j_n})\Big\rangle_B.
\end{align}
The sum $\sum_P$ represents the summation over all the transpositions $P(1,\dots, n)=j_1,\dots, j_n$. 
Replacing the integration indices $\tau_{j_1},\dots, \tau_{j_n} \rightarrow \tau_1,\dots, \tau_n$ for each separate term (the total number of such terms is $n!$) in the expression we obtain 
\begin{align}
\Big\langle T_\tau\Big\{e^{-\int_0^\beta d\tau \widetilde{H}_{SB}(\tau)}\Big\}\Big\rangle_B
=&\prod_k\sum_{n=0}^\infty \frac{1}{n!}
\prod_{j=1}^n \int_0^\beta\int_0^\beta d\tau_jd\tau'_j \Big\langle T_\tau b_k^\dag(\tau_j)b_k(\tau'_j)\Big\rangle_B 
\times \nonumber \\ \times&
T_\tau \Big\{f_k(\tau_1)\dots f_k(\tau_n)f_k(\tau'_1)\dots f_k(\tau'_m)\Big\}
=\nonumber \\
=&T_\tau \Big\{\exp\Big[\sum_k \int_0^\beta d\tau \int_0^\beta d\tau' G(\Omega_k,\tau-\tau')f_k(\tau)f_k(\tau')\Big]\Big\}.
\end{align} 
Here we introduced a thermal boson Green's function 
$\langle T_\tau b_k^\dag(\tau)b_k(\tau')\rangle_B=G(\Omega_k,\tau-\tau')$, where
\begin{equation}
\label{eq:Green_function}
G(\xi,\tau-\tau')= 
\frac{e^{(\tau-\tau')\xi}}{e^{\beta\xi}-1}\theta(\tau-\tau')+
\frac{e^{(\tau-\tau')\xi}}{1-e^{-\beta\xi}}\theta(\tau'-\tau).
\end{equation}
Introducing the spectral density function $\mathcal{I}(\xi)=\sum_k \lambda_k^2\delta(\xi-\Omega_k)$
one can write 
\begin{equation}
\Big\langle T_\tau\Big\{e^{-\int_0^\beta d\tau \widetilde{H}_{SB}(\tau)}\Big\}\Big\rangle_B=
T_\tau \Big\{\exp\Big[\int_0^\infty d\xi\,\mathcal{I}(\xi) \int_0^\beta d\tau \int_0^\beta d\tau' G(\xi,\tau-\tau')
F(\tau)F(\tau')\Big]\Big\}, 
\end{equation}
with $f_k(\tau)=\lambda_k F(\tau)=\lambda_k\sum_{j=1}^{M+1}\boldsymbol{\sigma}_j(\tau)\cdot\boldsymbol{n}_j$, 
see Eq.~(\ref{eq:F_tau}) for details.
Therefore the leading term of the aforementioned expression is
\begin{align}
\Big\langle T_\tau\Big\{e^{-\int_0^\beta d\tau \widetilde{H}_{SB}(\tau)}\Big\}\Big\rangle_B\approx& 1+
\int_0^\infty d\xi\,\mathcal{I}(\xi)\, T_\tau\Big\{\int_0^\beta d\tau \int_0^\beta d\tau' G(\xi,\tau-\tau')
F(\tau)F(\tau')\Big\}=\nonumber \\
=&1+\int_0^\infty d\xi\,\mathcal{I}(\xi)\int_0^\beta d\tau \int_0^\tau d\tau' 
\phi(\xi,\tau-\tau')F(\tau)F(\tau'), 
\end{align}
where we introduced the notation 
\begin{equation}
\label{eq:phi_function}
\phi(\xi,\tau-\tau')=\frac{e^{(\tau-\tau')\xi}}{e^{\beta\xi}-1} +
\frac{e^{-(\tau-\tau')\xi}}{1-e^{-\beta\xi}}.
\end{equation}
Remembering that the term $F(\tau)$ is a linear combination of the spin operators, 
we conclude that the leading term of the decomposition $F(\tau)F(\tau')$ 
describes the spin-spin correlation between the spin degrees of freedom in spin subsystem 
induced by its interaction with bosonic bath. 

\section{Average value of $\sigma_{M+1}^z$ operator for the case of $M+1$ spins}
\label{app:sigma_z}

The expression for the average value of the $\sigma_{M+1}^z$ operator 
up to the second order of the perturbation theory in parameters $f_1,f_2$ 
has the following form  
\begin{align}
\label{eq:sigma_z_full}
\langle\sigma_{M+1}^z\rangle\approx&\langle\sigma_{M+1}^z\rangle_S
\Big(1-\int_0^\infty d\xi\,\mathcal{I}(\xi)\int_0^\beta d\tau \int_0^\tau d\tau' 
\phi(\xi,\tau-\tau')\langle F(\tau)F(\tau')\rangle_S\Big)+ \nonumber \\ +& 
\int_0^\infty d\xi\,\mathcal{I}(\xi)\int_0^\beta d\tau \int_0^\tau d\tau' 
\phi(\xi,\tau-\tau')\langle F(\tau)F(\tau')\sigma_{M+1}^z\rangle_S,
\end{align}
where $\phi(\xi, \tau-\tau')$ is defined by Eq.~(\ref{eq:phi_function}) and 
$F(\tau)=\sum_{j=1}^{M+1}\boldsymbol{\sigma}_j(\tau)\cdot\boldsymbol{n}_j$, see Eq.~(\ref{eq:F_tau}).
Evaluating the integrands 
\begin{align}
\langle F(\tau)F(\tau')\rangle_S\langle\sigma_{M+1}^z\rangle_S=
&\Big[f_1^2\sum_{j,l=1}^M \langle\sigma_j^z\sigma_l^z\rangle_S+
f_2^2\cosh(\omega_{M+1}(\tau-\tau'))\Big]\langle\sigma_{M+1}^z\rangle_S+\nonumber \\&
+
f_2^2\sinh(\omega_{M+1}(\tau-\tau'))\langle\sigma_{M+1}^z\rangle_S^2,
\\
%=\nonumber \\ =&
%f_1^2\Big[M+2\sum_{j<l}\tanh\Big(\frac{\beta\omega_j}{2}\Big)\tanh\Big(\frac{\beta\omega_l}{2}\Big)\Big]+\nonumber \\+&
%f_2^2\cosh(\omega_{M+1}(\tau-\tau'))-f_2^2\sinh(\omega_{M+1}(\tau-\tau'))\tanh\Big(\frac{\beta\omega_{M+1}}{2}\Big),
\langle F(\tau)F(\tau')\sigma_{M+1}^z\rangle_S=&
f_1^2\sum_{j,l=1}^M\langle\sigma_j^z\sigma_l^z\rangle_S
\langle\sigma_{M+1}^z\rangle_S+f_2^2\cosh(\omega_{M+1}(\tau-\tau'))
\langle\sigma_{M+1}^z\rangle_S+ \nonumber \\&+f_2^2\sinh(\omega_{M+1}(\tau-\tau')),
%=\nonumber \\=&-
%f_1^2\tanh\left(\frac{\beta\omega_{M+1}}{2}\right)\Big[M+2\sum_{j<l}\tanh\Big(\frac{\beta\omega_j}{2}\Big)
%\tanh\Big(\frac{\beta\omega_l}{2}\Big)\Big]- \nonumber \\&-
%f_2^2\tanh\left(\frac{\beta\omega_{M+1}}{2}\right)\cosh(\omega_{M+1}(\tau-\tau'))+f_2^2\sinh(\omega_{M+1}(\tau-\tau')),
\end{align}
and substituting them into the Eq.~(\ref{eq:sigma_z_full}) we get 
\begin{align}
\langle\sigma_{M+1}^z\rangle\approx&\langle\sigma_{M+1}^z\rangle_S+
f_2^2 \Big[1-\langle\sigma_{M+1}^z\rangle_S^2\Big]\times \nonumber \\&\times
\int_0^\infty d\xi\,\mathcal{I}(\xi)\int_0^\beta d\tau \int_0^\tau d\tau' 
\phi(\xi,\tau-\tau')\sinh(\omega_{M+1}(\tau-\tau')),
\end{align}
where $\langle\sigma_{M+1}^z\rangle_S=-\tanh(\beta\omega_{M+1}/2)$ is the average value of the 
$\sigma_{M+1}^z$ spin operator in the absence of the spin-bath coupling. Therefore the value 
$\delta \langle\sigma_{M+1}^z\rangle\equiv \langle\sigma_{M+1}^z\rangle-\langle\sigma_{M+1}^z\rangle_S$, 
describes the correction to the $z$ component of the $(M+1)$th spin induced by the spin-bath interaction. 
The final result reads 
\begin{align}
\delta\langle&\sigma_{M+1}^z\rangle=f_2^2
\int_0^\infty d\xi\frac{\mathcal{I}(\xi)}{\left(\xi^2-\omega_{M+1}^2\right)^2  
} \times \nonumber \\ &\times 
\Big\{\frac{\coth \left(\frac{\beta\xi}{2}\right)}{\cosh^2\Big(\frac{\beta\omega_{M+1}}{2}\Big)} 
\left[\beta\omega_{M+1}\left(\xi ^2-\omega_{M+1}^2\right)+
\left(\xi^2+\omega_{M+1}^2\right)\sinh(\beta\omega_{M+1})\right]-
4\xi\omega_{M+1}\Big\}.
\end{align}

\section{Details of the calculation of $\sigma^2$, Eq.~(\ref{eq:sigma_definition})}
\label{app:intermediate}

In the leading order of the perturbation theory the Eq.~(\ref{eq:sigma_definition}) reads 
$\sigma^2\approx\langle\sigma_{M+1}^x\sigma_{M+2}^x\rangle$. Then one obtains
\begin{align}
\label{eq:sigma_squared}
\sigma^2\approx &\int_0^\infty d\xi\,\mathcal{I}(\xi)\int_0^\beta d\tau \int_0^\tau d\tau' \phi(\xi,\tau-\tau')
\langle F(\tau)F(\tau')\sigma_{M+1}^x\sigma_{M+2}^x\rangle_S=\nonumber \\
=&f_2^2\int_0^\infty d\xi\,
\frac{\mathcal{I}(\xi)\mathcal{K}(\xi)}{\left(\xi^2-\omega_{M+1}^2\right)\left(\xi^2-\omega_{M+2}^2\right)},
\end{align}
where $\phi(\xi,\tau-\tau')$ is defined by Eq.~(\ref{eq:phi_function}),  
$F(\tau)=\sum_{j=1}^{M+2}\boldsymbol{\sigma}_j(\tau)\cdot\boldsymbol{n}_j$. 
The function $\mathcal{K}(\xi)$ reads
\begin{align}
\mathcal{K}(\xi)=&a(\xi)+\tanh\Big(\frac{\beta\omega_{M+1}}{2}\Big)\tanh\Big(\frac{\beta\omega_{M+2}}{2}\Big)b(\xi)-
\nonumber \\-&
\tanh\Big(\frac{\beta\omega_{M+2}}{2}\Big)c(\xi)-\tanh\Big(\frac{\beta\omega_{M+1}}{2}\Big)d(\xi),
\end{align}
where we introduced  
\begin{align}
a(\xi)=&\xi\omega_{M+1}\sinh (\beta\omega_{M+1}) 
\left(\cosh(\beta\omega_{M+2})
\frac{\left(2\xi^2-\omega_{M+1}^2-\omega_{M+2}^2\right)}{\omega_{M+1}^2-\omega_{M+2}^2}-1\right)-\nonumber \\
-&\xi\omega_{M+2}\sinh(\beta\omega_{M+2})
\left(\cosh(\beta\omega_{M+1})\frac{\left(2\xi^2-\omega_{M+1}^2-\omega_{M+2}^2\right)}{\omega_{M+1}^2-\omega_{M+2}^2}+
1\right)+\nonumber \\+&
\omega_{M+1}\omega_{M+2}\coth \left(\frac{\beta\xi}{2}\right)\sinh(\beta\omega_{M+1})\sinh(\beta\omega_{M+2})
- \nonumber \\-&
4\xi^2 \coth \left(\frac{\beta\xi}{2}\right) \sinh^2\left(\frac{\beta\omega_{M+1}}{2}\right)\sinh^2\left(\frac{\beta\omega_{M+2}}{2}\right),
\end{align}
\begin{align}
b(\xi)=&\xi\omega_{M+2}\sinh(\beta\omega_{M+1}) \left(1-\frac{\cosh (\beta\omega_{M+2})
\left(2\xi^2-\omega_{M+1}^2-\omega_{M+2}^2\right)}{\omega_{M+1}^2-\omega_{M+2}^2}\right)+\nonumber \\+&
\xi\omega_{M+1}\sinh(\beta\omega_{M+2})\left(\frac{\cosh(\beta\omega_{M+1})\left(2\xi^2-\omega_{M+1}^2
-\omega_{M+2}^2\right)}{\omega_{M+1}^2-\omega_{M+2}^2}+1\right)
+\nonumber \\+&
4\omega_{M+1}\omega_{M+2}\coth\left(\frac{\beta\xi}{2}\right)\sinh^2\left(\frac{\beta\omega_{M+1}}{2}\right)\sinh^2
\left(\frac{\beta\omega_{M+2}}{2}\right)
-\nonumber \\-&
\xi^2\coth\left(\frac{\beta\xi}{2}\right)\sinh(\beta\omega_{M+1})\sinh(\beta\omega_{M+2}).
\end{align}
\begin{align}
c(\xi)=&\xi\omega_{M+1}\sinh(\beta\omega_{M+1})\sinh(\beta\omega_{M+2})\frac{\left(2\xi^2-\omega_{M+1}^2
-\omega_{M+2}^2\right)}{\omega_{M+1}^2-\omega_{M+2}^2}-
\nonumber \\-&
\xi\omega_{M+2}\Big[\cosh(\beta\omega_{M+1})\cosh(\beta\omega_{M+2})-1\Big]\frac{\left(2\xi^2-\omega_{M+1}^2-
\omega_{M+2}^2\right)}{\omega_{M+1}^2-\omega_{M+2}^2}+\nonumber \\+&
2\omega_{M+1}\omega_{M+2}\coth\left(\frac{\beta\xi}{2}\right)\sinh(\beta\omega_{M+1})
\sinh^2\left(\frac{\beta\omega_{M+2}}{2}\right)-\nonumber \\-&
2\xi^2\coth\left(\frac{\beta\xi}{2}\right)\sinh^2\left(\frac{\beta\omega_{M+1}}{2}\right) 
\sinh (\beta\omega_{M+2})
+\nonumber \\+&
\xi\omega_{M+2}\Big[\cosh(\beta\omega_{M+1})-\cosh(\beta\omega_{M+2})\Big],
\end{align}
\begin{align}
d(\xi)=&-\xi\omega_{M+2}\sinh(\beta\omega_{M+1})\sinh(\beta\omega_{M+2})\frac{\left(2\xi^2-\omega_{M+1}^2-
\omega_{M+2}^2\right)}{\omega_{M+1}^2-\omega_{M+2}^2}+ \nonumber \\ +&
\xi\omega_{M+1}\Big[\cosh(\beta\omega_{M+1})\cosh(\beta\omega_{M+2})-1\Big]\frac{\left(2\xi^2-\omega_{M+1}^2-\omega_{M+2}^2\right)}{\omega_{M+1}^2-\omega_{M+2}^2}+\nonumber \\ +&
2\omega_{M+1}\omega_{M+2}\coth\left(\frac{\beta\xi}{2}\right)\sinh^2\left(\frac{\beta\omega_{M+1}}{2}\right)
\sinh(\beta\omega_{M+2})
-\nonumber \\-&
2\xi^2\coth\left(\frac{\beta\xi}{2}\right)\sinh(\beta\omega_{M+1})\sinh^2\left(\frac{\beta\omega_{M+2}}{2}\right) 
+\nonumber  \\+&
\xi\omega_{M+1}\Big[\cosh(\beta\omega_{M+2})-\cosh(\beta\omega_{M+1})\Big].
\end{align}
After simplification one gets the following expression
\begin{align}
\mathcal{K}(\xi)=4
\Big[&\omega_{M+1}\omega_{M+2}\coth\left(\frac{\beta\xi}{2}\right)
\tanh\left(\frac{\beta\omega_{M+1}}{2}\right)\tanh\left(\frac{\beta\omega_{M+2}}{2}\right) +
 \nonumber \\ &+
\frac{\xi\omega_{M+1} (\xi^2-\omega_{M+1}^2)\tanh\left(\frac{\beta\omega_{M+1}}{2}\right)-
\xi\omega_{M+2} (\xi^2-\omega_{M+2}^2)\tanh\left(\frac{\beta\omega_{M+2}}{2}\right)}
{\omega_{M+1}^2-\omega_{M+2}^2}\Big],
\end{align}
which after substituting it in the expression (\ref{eq:sigma_squared}) finally leads to 
Eqs.~(\ref{eq:sigma_correlation}) and (\ref{eq:G}) for $\sigma^2$.  

\section{Average value of $\sigma^x$ for the case of $M+N+1$ spins and two independent baths}
\label{app:sigma_x}

We calculate the average $\langle\sigma^x\rangle$ using the formula
\begin{align}
\langle\sigma^x\rangle=&\frac{1}{Z}\text{Tr}\Big[e^{-\beta H}\sigma^x\Big],
\end{align}
with the Hamiltonian (\ref{eq:full_hamiltonian}).
We rewrite this expression in the form
\begin{align}
\langle\sigma^x\rangle=&\frac{1}{Z}\text{Tr}\Big[e^{-\beta H}\sigma^x\Big]=
%\frac{1}{Z}\text{Tr}\Big[UU^\dag e^{-\beta H}UU^\dag\sigma^x\Big]=\nonumber \\=&
%\frac{1}{Z}\text{Tr}\Big[U^\dag e^{-\beta H}U U^\dag\sigma^xU\Big]=
\frac{1}{Z}\text{Tr}\Big[e^{-\beta U^\dag HU} U^\dag\sigma^xU\Big],
\end{align}
where $U$ is a unitary operator and $Z=\text{Tr}[e^{-\beta H}]=\text{Tr}[e^{-\beta U^\dag HU}]$. 
We choose the unitary operator in the form of the Lang-Firsov type 
\begin{align}
U=&\exp\Big[-f_1\sum_{j=1}^M\sum_k\frac{\lambda_k}{\Omega_k}(b_k^\dag-b_k)\sigma_j^z\Big]\times
\exp\Big[-g_1\sum_{j=M+1}^{M+N}\sum_k\frac{\kappa_k}{\Upsilon_k}(d_k^\dag-d_k)\sigma_j^z\Big]. 
\end{align} 
Therefore $U^\dag\sigma^xU=\sigma^x$ and the transformed Hamiltonian 
$\mathcal{H}=U^\dag HU$
\begin{equation}
\mathcal{H}=\mathcal{H}_{S1}+\mathcal{H}_{S2}+\mathcal{H}_B+
\mathcal{H}_S+\mathcal{H}_{\text{int},S}+\mathcal{H}_{\text{int},S12}, 
\end{equation}
where 
\begin{align}
\mathcal{H}_{S1}=\frac{\omega_1}{2}\sum_{j=1}^M\sigma_j^z-\Omega f_1^2 \sum_{j,l=1}^M \sigma_j^z\sigma_l^z, \quad 
\mathcal{H}_{S2}=\frac{\omega_2}{2}\sum_{j=M+1}^{M+N}\sigma_j^z-
\Upsilon g_1^2 \sum_{j,l=M+1}^{M+N} \sigma_j^z\sigma_l^z,
\end{align}
are the new spin Hamiltonians, corresponding to the group of spins coupled to the first ($\mathcal{H}_{S1}$) and the second ($\mathcal{H}_{S2}$) bosonic baths. Here we introduced the reorganization energies 
$\Omega=\sum_k \lambda_k^2/\Omega_k$, $\Upsilon=\sum_k \kappa_k^2/\Upsilon_k$ of the first and the second bath, respectively. The Hamiltonian of the corresponding baths is
\begin{align}
\mathcal{H}_B=\sum_k (\Omega_k b_k^\dag b_k+\Upsilon_k d_k^\dag d_k).
\end{align}
Then,  
\begin{equation}
\mathcal{H}_S=\frac{\omega}{2}\sigma^z,
\end{equation}
is the Hamiltonian of the remaining (output) spin coupled to both bathes by term 
\begin{align}
\mathcal{H}_{\text{int},S}=&f_2\sigma^x \sum_k \lambda_k(b_k^\dag+b_k)+
g_2\sigma^x \sum_k \kappa_k(d_k^\dag+d_k).
\end{align}
Finally, the term 
\begin{align}
\mathcal{H}_{\text{int},S12}=&-2f_1f_2\Omega \sigma^x\sum_{j=1}^M\sigma_j^z
-2g_1g_2\Upsilon \sigma^x\sum_{j=M+1}^{M+N}\sigma_j^z, 
\end{align}
describes effective coupling between $M+N$ input spins $\sigma_j^z$ and the output spin $\sigma^x$.

Using the structure the spin Hamiltonians $\mathcal{H}_{S1}$ and $\mathcal{H}_{S2}$ one can calculate their eigenstates and eigenvalues. Namely, the eigenstates can be written as a product of all possible combinations of the  spin-up $|\uparrow\rangle$ and spin-down $|\downarrow\rangle$ states. 
For example, for the Hamiltonian $\mathcal{H}_{S1}$ has $2^M$ eigenstates 
\begin{equation}
|s_1,s_2,\dots s_M\rangle=|s_1\rangle|s_2\rangle\dots |s_M\rangle, 
\end{equation}
where $s_j=\uparrow, \downarrow$ for each $j=1,2,\dots M$. 
The eigenvalues of the Hamiltonian don't depend on the position of the spin-up and spin-down 
states in the eigenstate in $|s_1,s_2,\dots s_M\rangle$, 
but only on the total number of spin-down and spin-up in the corresponding eigenstate.
Therefore all the eigenstates with $m$ spin-down and $M-m$ spin-up states are characterized by the 
eigenvalue 
\begin{equation}
E_{1,m}=\frac{\omega_1}{2}(M-2m)-\Omega f_1^2(M-2m)^2,
\end{equation}    
which represents ${M \choose m}=M!/m!(M-m)!$ times degenerated energy level. 
The analogous consideration can be performed for the
Hamiltonian $\mathcal{H}_{S2}$.  
Note that the Hamiltonians $\mathcal{H}_{S1}$ and $\mathcal{H}_{S2}$ commute with the rest of the 
terms in the $\mathcal{H}$ and therefore, we can take a trace over the input spin states
\begin{align}
\text{Tr}_{S1}\Big[\text{Tr}_{S2}\Big[e^{-\beta \mathcal{H}}\Big]\Big]=&\sum_{m=0}^M\sum_{n=0}^N 
{M \choose m}{N \choose n}
\exp\Big\{-\beta\big[\frac{\omega_1}{2}(M-2m)+
\frac{\omega_2}{2}(N-2n)\big]\Big\}
\times \nonumber \\ \times &
\exp\Big\{\beta\big[\Omega f_1^2(M-2m)^2+
\Upsilon g_1^2(N-2n)^2\big]\Big\}
\times \nonumber \\ \times &
\exp\Big\{-\beta\big[\mathcal{H}_B+\mathcal{H}_{\text{int},S}+
\big(\mathcal{H}_S-\frac{\omega_{mn}}{2}\sigma^x\big)\big]\Big\}, 
\end{align}
where $\omega_{mn}\equiv 4f_1f_2\Omega(M-2m)+4g_1g_2\Upsilon(N-2n)$.
We transform the last exponent in the expression to the form
\begin{align}
\exp\Big\{-\beta&\big[\mathcal{H}_{mn}+\mathcal{H}_{\text{int},S}\big]\Big\}=\exp\Big\{-\beta\mathcal{H}_{mn}\Big\}
T_\tau\Big\{\exp\Big(-\int_0^\beta d\tau \widetilde{H}_{\text{int},S}(\tau)\Big)\Big\}
\end{align}
using the notations 
$\mathcal{H}_{mn}=\mathcal{H}_B+\big(\mathcal{H}_S-\frac{\omega_{mn}}{2}\sigma^x\big)$ 
and
\begin{align}
\widetilde{H}_{\text{int},S}(\tau)=&\exp\Big\{\tau\mathcal{H}_{mn}\Big\} H_{\text{int},S}
\exp\Big\{-\tau\mathcal{H}_{mn}\Big\}%\nonumber \\ %\nonumber \\=&
%f_2\sigma^x(\tau)\sum_k \lambda_k[b_k^\dag(\tau)+b_k(\tau)]+g_2\sigma^x(\tau)\sum_k\kappa_k[d_k^\dag(\tau)+d_k(\tau)]
=\nonumber \\=&
\sum_k f_k(\tau)[b_k^\dag(\tau)+b_k(\tau)]+\sum_k g_k(\tau)[d_k^\dag(\tau)+d_k(\tau)].
\end{align}
Here $f_k(\tau)\equiv f_2\lambda_k\sigma^x(\tau)$, $g_k(\tau)\equiv g_2\kappa_k\sigma^x(\tau)$,
$b_k(\tau)=e^{-\Omega_k\tau}b_k$, $b_k^\dag(\tau)=e^{\Omega_k\tau}b_k^\dag$, 
$d_k(\tau)=e^{-\Upsilon_k\tau}d_k$, $d_k^\dag(\tau)=e^{\Upsilon_k\tau}d_k^\dag$,
\begin{align}
\sigma^x(\tau)=&[\sin^2\theta_{mn}+\cos^2\theta_{mn}\cosh(\tau R_{mn})]\sigma^x+
i\cos\theta_{mn}\sinh(\tau R_{mn})\sigma^y+ \nonumber \\+&
\sin\theta_{mn}\cos\theta_{mn}[\cosh(\tau R_{mn})-1]\sigma^z,
%\sigma^y(\tau)=&-i\cos\theta_{mn}\sinh(\tau R_{mn})\sigma^x+\cosh(\tau R_{mn})\sigma^y-
%i\sin\theta_{mn}\sinh(\tau R_{mn})\sigma^z,\\
%\sigma^z(\tau)=&\sin\theta_{mn}\cos\theta_{mn}[\cosh(\tau R_{mn})-1]\sigma^x+
%i\sin\theta_{mn}\sinh(\tau R_{mn})\sigma^y+ 
%[\cos^2\theta_{mn}+\sin^2\theta_{mn} \cosh(\tau R_{mn})]\sigma^z,
\end{align}
with $\theta_{mn}\equiv\arccos(\omega/R_{mn})$, $R_{mn}=\sqrt{\omega^2+\omega_{mn}^2}$.
Now we trace out the bosonic degrees of freedom from the reduced density operator. To do it we introduce the 
average
\begin{align}
\langle \star\rangle_B=\frac{1}{Z_B}\text{Tr}_B\Big[e^{-\beta \mathcal{H}_B}\star\Big], 
\end{align}  
where $Z_B=\text{Tr}_B\Big[e^{-\beta \mathcal{H}_B}\Big]$.
Then, the $\sigma$-dependent reduced density matrix is 
\begin{equation}
\rho_S=\frac{Z_B}{Z}\Big\langle\text{Tr}_{S1}\Big[\text{Tr}_{S2}\Big[e^{-\beta \mathcal{H}}\Big]\Big]\Big\rangle_B,
\end{equation}
where $Z$ is the partition function of the full system.   
After the averaging procedure we obtain 
\begin{align}
\rho_S=&\frac{Z_B}{Z}
\sum_{m=0}^M\sum_{n=0}^N {M \choose m}{N \choose n}
\exp\Big\{-\beta\big[\frac{\omega_1}{2}(M-2m)+
\frac{\omega_2}{2}(N-2n)\big]\Big\}\times \nonumber \\ \times& 
\exp\Big\{\beta\big[\Omega f_1^2(M-2m)^2+
\Upsilon g_1^2(N-2n)^2\big]\Big\}
\exp\Big\{-\beta\big(\mathcal{H}_S-\frac{\omega_{mn}}{2}\sigma^x\big)\Big\}
\times \nonumber \\ \times &
\Big\langle T_\tau\Big\{\exp\Big(-\int_0^\beta d\tau 
\widetilde{H}_{\text{int},S}(\tau)\Big)\Big\}\Big\rangle_B.
\end{align}
Then we calculate the average of the $T_\tau$-exponents using the results of
Appendix~\ref{app:derivation} 
\begin{align}
\label{eq:reduced}
\rho_S=&\frac{Z_B}{Z}
\sum_{m=0}^M\sum_{n=0}^N {M \choose m}{N \choose n}
\exp\Big\{-\beta\big[\frac{\omega_1}{2}(M-2m)+
\frac{\omega_2}{2}(N-2n)\big]\Big\}
\times \nonumber \\ \times &
\exp\Big\{\beta\big[\Omega f_1^2(M-2m)^2+\Upsilon g_1^2(N-2n)^2\big]\Big\}
\exp\Big\{-\beta\big(\mathcal{H}_S-\frac{\omega_{mn}}{2}\sigma^x\big)\Big\}
\times \nonumber \\ \times &
T_\tau \exp\Big\{f_2^2\int_0^\infty d\xi\,\mathcal{I}_1(\xi) \int_0^\beta d\tau \int_0^\beta d\tau' G(\xi,\tau-\tau')
\sigma^x(\tau)\sigma^x(\tau')\Big\}
\times \nonumber \\ \times &
T_\tau \exp\Big\{g_2^2\int_0^\infty d\xi\,\mathcal{I}_2(\xi) \int_0^\beta d\tau \int_0^\beta d\tau' G(\xi,\tau-\tau')
\sigma^x(\tau)\sigma^x(\tau')\Big\},
\end{align}
with the Green's function (\ref{eq:Green_function}) and    
and spectral density functions (\ref{eq:spectral_density_functions}). 
In order to reduce this expression for the most convenient form for further calculations we use the identity 
\begin{align}
\exp\Big\{&-\beta \big(\frac{\omega}{2}\sigma^z-\frac{\omega_{mn}}{2}\sigma^x\big)\Big\}=
\exp\Big\{i\frac{\theta_{mn}}{2}\sigma^y\Big\}
\exp\Big\{-\beta\frac{R_{mn}}{2}\sigma^z\Big\}
\exp\Big\{-i\frac{\theta_{mn}}{2}\sigma^y\Big\}
\end{align}     
and we rewrite $\rho_S$ in the form 
\begin{align}
\rho_S=&\exp\Big\{i\frac{\theta_{mn}}{2}\sigma^y\Big\}\Big[\frac{Z_B}{Z}
\sum_{m=0}^M\sum_{n=0}^N {M \choose m}{N \choose n}
\exp\Big\{-\beta\big[\frac{\omega_1}{2}(M-2m)+
\frac{\omega_2}{2}(N-2n)\big]\Big\}
\times \nonumber \\ \times &
\exp\Big\{\beta\big[\Omega f_1^2(M-2m)^2+\Upsilon g_1^2(N-2n)^2\big]\Big\}
\exp\Big\{-\beta\frac{R_{mn}}{2}\sigma^z\Big\}
\times \nonumber \\ \times &
T_\tau \exp\Big\{f_2^2\int_0^\infty d\xi\,\mathcal{I}_1(\xi) \int_0^\beta d\tau \int_0^\beta d\tau' G(\xi,\tau-\tau')
\Sigma^x(\tau)\Sigma^x(\tau')\Big\}
\times \nonumber \\ \times &
 T_\tau\exp\Big\{g_2^2\int_0^\infty d\xi\,\mathcal{I}_2(\xi) \int_0^\beta d\tau \int_0^\beta d\tau' G(\xi,\tau-\tau')
\Sigma^x(\tau)\Sigma^x(\tau')\Big\}\Big] \exp\Big\{-i\frac{\theta_{mn}}{2}\sigma^y\Big\}.
\end{align}
Here we introduced 
\begin{align}
\Sigma^x(\tau)=&e^{-i\frac{\theta_{mn}}{2}\sigma^y}\sigma^x(\tau)
e^{i\frac{\theta_{mn}}{2}\sigma^y}=
\cos\theta_{mn}[\cosh(R_{mn}\tau)\sigma^x+i\sinh(R_{mn}\tau)\sigma^y]-\sin\theta_{mn}\sigma^z
\nonumber \\ \equiv &
\mathbf{N}(\tau)\cdot\boldsymbol{\sigma},
\end{align}	
which gives the following expression 
\begin{align}
\Sigma^x(\tau)\Sigma^x(\tau')=&\mathbf{N}(\tau)\cdot\mathbf{N}(\tau')+i\boldsymbol{\sigma}\cdot(\mathbf{N}(\tau)\times\mathbf{N}(\tau'))=\nonumber \\ =&(\cos^2\theta_{mn}\cosh[R_{mn}(\tau-\tau')]+\sin^2\theta_{mn})\mathcal{I}
+
\sigma^z\cos^2\theta_{mn}\sinh[R_{nm}(\tau-\tau')]+\nonumber \\+&
\sigma^x\sin\theta_{mn}\cos\theta_{mn}[\sinh(R_{nm}\tau)-\sinh(R_{nm}\tau')]+\nonumber \\+& 
i\sigma^y\sin\theta_{mn}\cos\theta_{mn}[\cosh(R_{nm}\tau)-\cosh(R_{nm}\tau')]
\end{align}	 		
Taking into account the identity $\text{Tr}_S[\rho_S]=1$ and cyclic properties of trace operation we obtain 
\begin{align}
\label{eq:final_Z}
Z=&Z_B
\sum_{m=0}^M\sum_{n=0}^N {M \choose m}{N \choose n}
\exp\Big\{-\beta\big[\frac{\omega_1}{2}(M-2m)+\frac{\omega_2}{2}(N-2n)\big]\Big\}
\times \nonumber \\ \times &
\exp\Big\{\beta\big[\Omega f_1^2(M-2m)^2+\Upsilon g_1^2(N-2n)^2\big]\Big\}
\text{Tr}_S\Big[\exp\Big\{-\beta\frac{R_{mn}}{2}\sigma^z\Big\}
\times \nonumber \\ \times &
T_\tau \exp\Big\{f_2^2\int_0^\infty d\xi\,\mathcal{I}_1(\xi) \int_0^\beta d\tau \int_0^\beta d\tau' G(\xi,\tau-\tau')
\Sigma^x(\tau)\Sigma^x(\tau')\Big\}
\times \nonumber \\ \times &
T_\tau \exp\Big\{g_2^2\int_0^\infty d\xi\,\mathcal{I}_2(\xi) \int_0^\beta d\tau \int_0^\beta d\tau' G(\xi,\tau-\tau')
\Sigma^x(\tau)\Sigma^x(\tau')\Big\}\Big].
\end{align} 	
Therefore the expression for the coherence takes the form 
\begin{align}
\label{eq:final_expression}
\langle\sigma^x\rangle=&\frac{Z_B}{Z}
\sum_{m=0}^M\sum_{n=0}^N {M \choose m}{N \choose n}
\exp\Big\{-\beta\big[\frac{\omega_1}{2}(M-2m)+\frac{\omega_2}{2}(N-2n)\big]\Big\}
\times \nonumber \\ \times & 
\exp\Big\{\beta\big[\Omega f_1^2(M-2m)^2+\Upsilon g_1^2(N-2n)^2\big]\Big\}
\text{Tr}_S\Big[
\exp\Big\{-\beta\frac{R_{mn}}{2}\sigma^z\Big\}
\times \nonumber \\ \times &
T_\tau \exp\Big\{f_2^2\int_0^\infty d\xi\,\mathcal{I}_1(\xi) \int_0^\beta d\tau \int_0^\beta d\tau' G(\xi,\tau-\tau')
\Sigma^x(\tau)\Sigma^x(\tau')\Big\}
\times \nonumber \\ \times &
T_\tau \exp\Big\{g_2^2\int_0^\infty d\xi\,\mathcal{I}_2(\xi) \int_0^\beta d\tau \int_0^\beta d\tau' G(\xi,\tau-\tau')
\Sigma^x(\tau)\Sigma^x(\tau')\Big\} \times \nonumber \\ \times&
(\cos\theta_{mn}\sigma^x-\sin\theta_{mn}\sigma^z)\Big].
\end{align}
Note that this is the exact result which takes into account non-perturbatively the coupling constants 
$f_1,f_2,g_1,g_2$ of the Hamiltonian. Moreover, the perturbative calculation of 
$T_\tau$ exponents in this expression requires $f_2,g_2\ll 1$, which is much better assumption than one used in  
Secs.~\ref{sec:synthesizing_coherence} and \ref{sec:coherence_multiplexing}, 
namely $f_2\ll 1/M$ with the large number $M$.
  
In order to complete the new scheme of the derivation of $\langle\sigma^x\rangle$, developed in this section, 
we provide the algorithm of calculation of the $T_\tau$-exponent in Eqs.~(\ref{eq:final_Z}) and 
(\ref{eq:final_expression}) perturbatively in the small coupling parameter $\kappa (=f_2\,\text{or}\,g_2)$ 
\begin{align}
T_\tau \exp\Big\{\kappa^2 & \int_0^\infty d\xi\,\mathcal{I}(\xi) \int_0^\beta d\tau \int_0^\beta d\tau' 
G(\xi,\tau-\tau')
\Sigma^x(\tau)\Sigma^x(\tau')\Big\}= \nonumber \\ =&
\sum_{n=0}^\infty \frac{\kappa^{2n}}{n!}\left[\prod_{j=1}^n 
\int_0^\infty d\xi_j\,\mathcal{I}(\xi_j) \int_0^\beta d\tau_{2j-1} \int_0^\beta d\tau_{2j}\, 
G(\xi_j,\tau_{2j-1}-\tau_{2j})\right]\times \nonumber \\ \times&  T_\tau\Big[\Sigma^x(\tau_1)\Sigma^x(\tau_2)\dots\Sigma^x(\tau_{2n})\Big] 
=\nonumber \\ =&
1+\kappa^2 \int_0^\infty d\xi_1\,\mathcal{I}(\xi_1) \int_0^\beta d\tau_1 \int_0^{\tau_1} d\tau_2\, 
\phi(\xi_1,\tau_1-\tau_2)
\Sigma^x(\tau_1)\Sigma^x(\tau_2) + \nonumber \\ +&
\kappa^4 \int_0^\infty d\xi_1\,\mathcal{I}(\xi_1) \int_0^\infty d\xi_2\,\mathcal{I}(\xi_2)  
\int_0^\beta d\tau_1 \int_0^{\tau_1} d\tau_2 \int_0^{\tau_2} d\tau_3 \int_0^{\tau_3} d\tau_4 \times 
\nonumber \\ \times& 
\Big[\phi(\xi_1,\tau_1-\tau_2)\phi(\xi_2,\tau_3-\tau_4)+\phi(\xi_1,\tau_1-\tau_3)\phi(\xi_2,\tau_2-\tau_4)+
\nonumber \\ &+
\phi(\xi_1,\tau_1-\tau_4)\phi(\xi_2,\tau_2-\tau_3)\Big]\,\,
\Sigma^x(\tau_1)\Sigma^x(\tau_2)\Sigma^x(\tau_3)\Sigma^x(\tau_4)+O(\kappa^6).
\end{align} 
Here $\mathcal{I}(\xi)$ represents the bosonic spectral density function, which in our case is 
$\mathcal{I}_1(\xi)$/$\mathcal{I}_2(\xi)$ for the left/right bath.  
Using the previously obtained result for the product of operators $\Sigma^x(\tau)\Sigma^x(\tau')$ one can derive
the leading correction term of the $T_\tau$-exponent, which is needed for the calculation of the coherence 
$\langle\sigma^x\rangle$ in higher orders in the coupling parameters $f_2,g_2$  
\begin{align}
\label{eq:leading_order}
\int_0^\beta d\tau_1 \int_0^{\tau_1} d\tau_2\, \phi(\xi,\tau_1-\tau_2)\Sigma^x(\tau)\Sigma^x(\tau')=
A\mathcal{I}+B\sigma^z+C\sigma^x+D\sigma^y,
\end{align}
where
\begin{align}
A=&\cos^2\theta_{mn}\frac{\beta\xi\left(\xi^2-R_{mn}^2\right)+2\left(\xi^2+R_{mn}^2\right) 
\coth\left(\frac{\beta\xi}{2}\right) \sinh^2\left(\frac{\beta R_{mn}}{2}\right)}{(\xi^2-R_{mn}^2)^2}-\nonumber \\-& 
\cos^2\theta_{mn}\frac{
2\xi R_{mn}\sinh(\beta  R_{mn})}{(\xi^2-R_{mn}^2)^2}+\sin^2\theta_{mn}\frac{\beta}{\xi}, \\
B=&\cos^2\theta_{mn}\frac{\left(\xi^2+R_{mn}^2\right)\coth\left(\frac{\beta\xi}{2}\right)\sinh(\beta R_{mn})
-\beta R_{mn}^3\coth \left(\frac{\beta\xi}{2}\right)}{\left(\xi^2-R_{mn}^2\right)^2}
+\nonumber \\+&
\cos^2\theta_{mn}\frac{\xi R_{mn}\left[\beta\xi\coth\left(\frac{\beta\xi}{2}\right)-
4\cosh^2\left(\frac{\beta R_{mn}}{2}\right)\right]}{\left(\xi^2-R_{mn}^2\right)^2}, \\
C=&2\sin\theta_{mn}\cos\theta_{mn}\frac{\xi\coth\left(\frac{\beta\xi}{2}\right)\sinh(\beta R_{mn})-2
R_{mn} \cosh^2\left(\frac{\beta R_{mn}}{2}\right)}{\xi(\xi^2-R^2_{mn})},\\
D=&2i\sin\theta_{mn}\cos\theta_{mn}\frac{2\xi\coth\left(\frac{\beta\xi}{2}\right)\sinh^2
\left(\frac{\beta  R_{mn}}{2}\right)-R_{mn}\sinh(\beta R_{mn})}{\xi\left(\xi^2-R_{mn}^2\right)}.
\end{align}
\end{widetext}
Integrating the result (\ref{eq:leading_order}) with the spectral density function $\mathcal{I}(\xi)$ over the
parameter $\xi$ one gets the expression for $T_\tau$-exponent up to quadratic terms of the parameter
$\kappa(=f_2\,\text{and}\,g_2)$ including. Substituting this result into Eqs.~(\ref{eq:final_Z}) and (\ref{eq:final_expression}), and then taking the trace $\text{Tr}_S[\cdots]$ over $\sigma$'s spin degrees of freedom 
one obtains next-to-leading-order expression in coupling parameters $f_2,g_2$ for the coherence.

\end{document}